\documentclass[12pt,psamsfonts,reqno]{amsart}

\theoremstyle{definition}

\theoremstyle{remark}

\newcommand{\vev}[1]{\left\langle #1 \right\rangle}
\newcommand{\ket}[1]{\left |  #1 \right \rangle}
\newcommand{\bra}[1]{\left \langle  #1 \right |}

\newcommand {\CalN} {\mathcal N}

\newcommand {\CalS} {\mathcal S}

\newcommand {\CalX} {\mathcal X}

\newcommand {\BI}   {\mathbb I}

\newcommand {\BR}   {\mathbb R}
\newcommand {\BZ}   {\mathbb Z}
\newcommand {\BC}   {\mathbb C}

\newcommand{\bb}{\mathbf{b}}
\newcommand{\bc}{\mathbf{c}}

\newcommand{\bV}{\mathbf{V}}

\newcommand{\bH}{\mathbf{H}}
\newcommand{\bM}{\mathbf{M}}
\newcommand{\bN}{\mathbf{N}}
\newcommand{\bK}{\mathbf{K}}
\newcommand{\bQ}{\mathbf{Q}}

\newcommand{\bX}{\mathbf{X}}

\newcommand{\bY}{\mathbf{Y}}
\newcommand{\by}{\mathbf{y}}

\newcommand{\msS}{\mathscr{S}}
\newcommand{\si}{\mathsf{i}}
\newcommand{\sn}{\mathsf{n}}

\newcommand{\sk}{\mathsf{k}}

\newcommand{\sA}{\mathsf{A}}

\newcommand{\sS}{\mathsf{S}}
\newcommand{\sV}{\mathsf{V}}

\newcommand{\sT}{\mathsf{T}}

\newcommand{\sY}{\mathsf{Y}}

\newcommand{\g}{\mathfrak{g}}

\newcommand{\frakM}{\mathfrak{M}}

\newcommand{\fq}{\mathfrak{q}}
\newcommand{\ep}{\epsilon}

\DeclareMathOperator{\ch}{ch}

\DeclareMathOperator{\rk} {rk}

\newcommand{\Res}[1]{\underset{#1}{\operatorname{Res}}}

\numberwithin{equation}{section}

\usepackage[margin=1in]{geometry}

\usepackage{amsmath}
\usepackage{amssymb}
\usepackage{amsxtra}
\usepackage{amscd}
\usepackage[mathscr]{euscript}
\usepackage{mdframed}

\usepackage{todonotes}

\usepackage[pagebackref=false]{hyperref}
\hypersetup{colorlinks = false, ,extension = notused,linkcolor = blue, anchorcolor = red,citecolor = blue,filecolor = red,pagecolor = red,urlcolor = blue}
\usepackage[numbers,sort&compress]{natbib}
\usepackage{hypernat}

\usepackage{iftex}
\ifxetex
        \usepackage{fontspec}
\else
\fi

\usepackage{float}
\usepackage{tikz-cd}
\usepackage[mathscr]{euscript}

\begin{document}

\title{Fractional quiver W-algebras}

\author{Taro Kimura}
\author{Vasily Pestun}

\address{Taro Kimura, Keio University, Japan}
\address{Vasily Pestun, IHES, France} 

\begin{abstract} 
 We introduce quiver gauge theory associated with the non-simply-laced type fractional quiver, and define fractional quiver W-algebras by using construction of~\cite{Kimura:2015rgi,Kimura:2016dys} with representation of fractional quivers.
\end{abstract}

\maketitle 

\tableofcontents

\parskip=4pt

 \section{Introduction}

Recently we proposed quiver gauge theoretic construction of $q$-deformed W-algebra~\cite{Kimura:2015rgi,Kimura:2016dys} through double quantum deformation of the geometric correspondence between 4d $\mathcal{N}=2$ (5d $\mathcal{N}=1$; 6d $\mathcal{N} = (1,0)$) gauge theory and the algebraic integrable systems~\cite{Gorsky:1995zq,Martinec:1995by,Donagi:1995cf,Nekrasov:2009rc,Nekrasov:2009zz}.
Our construction is orthogonal\footnote{The M-theory brane picture for A-series is rotated by 90 degrees.} to the AGT relation~\cite{Alday:2009aq,Wyllard:2009hg} and its $q$-deformed version~\cite{Awata:2009ur}. In contrast to the AGT relation, which associates $G$-Hitchin system to
a pure gauge theory with simple gauge group $G$, and after double $(\ep_1, \ep_2)$-quantization
one obtains $W_{\ep_1/\ep_2}(G)$-algebra,  in the quiver construction
$W(\Gamma)$-algebra comes from $\Gamma$-quiver gauge theory. The quiver W-algebra $W(\Gamma)$
can be interpreted as $\ep_2$-deformation of the ring of commuting Hamiltonians of the $\ep_1$-quantized integrable system~\cite{Nekrasov:2012xe,Nekrasov:2013xda} into an associative algebra of conserved currents of $q$-deformed 2d Toda field theory.
In our construction, the quiver $\Gamma$ is not necessarily required to be associated with the finite-type Dynkin diagram.

The $qq$-character~\cite{Nekrasov:2015wsu,
Nekrasov:2016qym,Nekrasov:2016ydq}  defines the generating current of the corresponding W-algebra.
This construction allows us to consider affine quiver theory, e.g. $\CalN=2^*$ theory ($\widehat{A}_0$ quiver), and define the W-algebra associated with affine Lie algebra.
In this case the bifundamental (adjoint) mass plays an essential role as a deformation parameter of W-algebra.

In the preceding papers~\cite{Kimura:2015rgi,Kimura:2016dys}, we have considered generic simply-laced quivers.
When the quiver diagram $\Gamma$ coincides with the Dynkin diagram of the finite Lie algebra, in particular, $\Gamma=ADE$, our construction reproduces Frenkel--Reshetikhin's definition of the $q$-deformed W-algebra~\cite{Frenkel:1996,Frenkel:1997,Frenkel:1998} and also \cite{Shiraishi:1995rp,Awata:1995zk}.
The aim of this paper is to extend our construction of quiver W-algebra to the non-simply-laced quiver.
For the non-simply-laced algebra, the root length can be different from each other in general, and is not invariant under the Langlands dual.
In the gauge theory, the Langlands dual exchanges the $\Omega$-background (equivariant) parameters $\epsilon_1 \leftrightarrow \epsilon_2$.
Thus the quiver gauge theory corresponding to the non-simply-laced algebra should depend on $\epsilon_1$ and $\epsilon_2$ in a different way.
In particular, its dependence could be different for the vector and hypermultiplets assigned to each node of quiver-Dynkin diagram.

In this paper we define the fractional quiver gauge theory, whose charge under the spacetime rotation depends on each quiver node. Let $\Gamma_0$ be the set of nodes of the quiver $\Gamma$ and $q_1, q_2 \in \BC^{\times}$ be
 the equivariant parameters of $\Omega$-background \cite{Moore:1997dj,Nekrasov:2002qd}. 
To every node $i \in \Gamma_0$ 
we assign a positive integer $d_i \in \BZ_{>0}$ and then declare
the equivariant parameters for fields at node $i$ to be 
$(q_1^{d_i}, q_2)$.
This construction is actually motivated by Frenkel--Reshetikhin's construction of the $q$-deformed W-algebra of non-simply-laced type~\cite{Frenkel:1997}, which is applicable to any simple Lie algebras.
We show that the charge $(d_i)_{i \in \Gamma_0}$ plays a role of the relative root length of the corresponding algebra.
At node $i$ under such assignment of charge there is  $\BZ_{d_i}$ symmetry $q_1 \to e^{2n\pi\iota/d_i} q_1$ with $n = 0,\ldots,d_i-1$, which is similar to the orbifold $(\BC/\BZ_{d_i}) \times \BC$ with the identification $(z_1, z_2) \sim ( e^{2\pi\iota/d_i} z_1, z_2)$, used to study the instanton moduli space in the presence of the surface operator~\cite{FFNR:2011SM,Finkelberg:2014JEMS,Kanno:2011fw}.
%We remark that the fractional quiver gauge theory does not reproduce the gauge theory on this orbifold, or rather, the root of unity limit $q_1 \to e^{2\pi\iota/d_i}$ would yield a proper result~\cite{Kimura:2011zf,Kimura:2011gq}.
A geometric realization of fractional quiver will be discussed in a forthcoming paper~\cite{Kimura:2018}.

Applying our construction to the fractional quiver gauge theory, we obtain W-algebras associated with non-simply-laced algebras, which reproduces the definition given by Frenkel--Reshetikhin~\cite{Frenkel:1996,Frenkel:1997}.
With generic quiver which does not correspond to any finite Lie algebras, our construction gives rise to non-simply-laced (twisted) affine and hyperbolic W-algebras, which we call {\em fractional quiver W-algebras} in general.
We also remark that there are several related works on non-simply-laced quiver gauge theory, especially, associated with finite-dimensional Lie algebras, with the little string theory perspective~\cite{Aganagic:2015cta,Aganagic:2017smx,Haouzi:2017vec}, and three-dimensional mirror symmetry~\cite{Cremonesi:2014xha,Dey:2016qqp}.

\subsection*{Acknowledgements}
The work of T.K. was supported in part by Keio Gijuku Academic Development Funds, JSPS Grant-in-Aid for Scientific Research (No.~JP17K18090), the MEXT-Supported Program for the Strategic Research Foundation at Private Universities ``Topological Science'' (No.~S1511006), JSPS Grant-in-Aid for Scientific Research on Innovative Areas ``Topological Materials Science'' (No.~JP15H05855), and ``Discrete Geometric Analysis for Materials Design'' (No.~JP17H06462). V.P. acknowledges grant  RFBR 16-02-01021. The research of V.P. on this project has received funding from the European Research Council (ERC) under the European Union's Horizon 2020 research and innovation program (QUASIFT grant agreement 677368).

\section{Fractional quiver gauge theory}

\subsection{Gauge theory definition}

We use the notations of~\cite{Kimura:2015rgi,Kimura:2016dys}.

Let $\Gamma$ be a quiver with the set of nodes (vertices) $\Gamma_0$ and the set of arrows (edges) $\Gamma_1$. An edge from $i$ to $j$ is denoted by $e: i \to j$.
A fractional quiver $(\Gamma,d)$ is a quiver $\Gamma$ decorated by positive
integer labels on the vertices $d: \Gamma_0 \to \BZ_{>0}$, so that to each vertex $i$ there is associated number $d_i > 0$. The meaning of the number $d_i$ is
the relative root length squared of the respective Lie algebra associated to the fractional quiver as will be clear later in (\ref{eq:symCartan_0th}).

We define $d$-fractional quiver theory on $\BC^2$ as follows.
We consider the ring $R = \BC[z_1, z_2]$ and 
in the node $i \in \Gamma_0$ we replace the ring $\BC[z_1, z_2]$ by the ring $R_i =\BC[z_1^{d_i}, z_2]$.
%which corresponds to the structure sheaf of the quotient space
%\begin{equation}
% (z_1,z_2) \sim ( z_1, z_2 e^{2\pi \imath/d_i})
%  \, .
%\end{equation}
The equivariant gauge theory counts $R_i$ ideals.
This construction is similar, but different from the instanton counting on the orbifold $(\BC / \BZ_{d_i}) \times \BC$ itself, which is used to implement the surface operator~\cite{FFNR:2011SM,Finkelberg:2014JEMS,Kanno:2011fw}.
See also~\cite{Kimura:2011zf,Kimura:2011gq} for a realization of the orbifold using the equivariant parameter.

Namely, for the observable sheaves over the instanton moduli space associated to the ring $R_i$, which is a pullback of the universal sheaves $(\bY_i)_{i \in \Gamma_0}$, we have
\begin{equation}
  [\bY_o]_i = [\mathbf{N}_i]  - [\Lambda \bQ_i] [\bK_i]
\end{equation}
where we denote by $o$ the $\sT$-fixed point in $\BC^2$ under the equivariant action, namely $(z_1,z_2) = (0,0)$.
The graded by nodes vector space $\bN = (\bN_i)_{i \in \Gamma_0}$ is the framing space for each node of quiver in the ADHM construction, and the graded by nodes vector space $\bK = (\bK_i)_{i \in \Gamma_0}$ is associated with the ideal generated by the partition $(\lambda_{i,\alpha})_{i \in \Gamma_0, \alpha = [1 \ldots \sn_i]}$, characterizing the equivariant $\sT$-fixed point of the moduli space, with $(\sn_i)_{i \in \Gamma_0}$ the rank of gauge group $U(\sn_i)$ assigned to the node $i \in \Gamma_0$.
The Chern characters of $\bN_i$ and $\bK_i$ are given by
\begin{equation}
 \ch \mathbf{N}_i = \sum_{\alpha=1}^{\sn_i} \nu_{i,\alpha}
  \, , \qquad
 \ch \mathbf{K}_i = \sum_{\alpha=1}^{\sn_i} \sum_{s \in \lambda_{i,\alpha}} \nu_{i,\alpha} q_1^{d_i (s_1 - 1)}
  q_2^{s_2 - 1}
\end{equation}
and $\ch \Lambda \bQ_i = (1 - q_1^{d_i})(1 - q_2)$.
The pair $(q_1, q_2)$ denotes the multiplicative equivariant parameters for the space-time rotation with $(q_1,q_2)=(e^{\epsilon_1},e^{\epsilon_2})$, and $(\nu_{i,\alpha})_{i\in\Gamma_0, \, \alpha\in[1\ldots\sn_i]}$ are the multiplicative Coulomb moduli parameters.
In this paper we use multiplicative (5d/K-theoretic) notation for
the equivariant parameters.
See~\cite{Kimura:2015rgi,Kimura:2016dys} for more details on the definition.

For a quiver $\Gamma$, we assign a vector multiplet to each node $i \in \Gamma_0$ and a hypermultiplet in bifundamental representation to each edge $e \in \Gamma_1$.
The (anti)fundamental hypermultiplet will be added separately (See Sec.~\ref{sec:fund_mat}).
A vector multiplet contribution in node $i$ comes from
\begin{equation}
 [\mathbf{V}_i] =
% -\frac{1}{ [\Lambda \bQ^{\vee}_{d_i}]} [\bY_o]_i [\bY_o]_i^{\vee}
% - \frac{1}{[\Lambda \bQ_i^\vee]}
 \frac{1}{[\Lambda \bQ_i]}  
 [\bY_o^\vee]_i [\bY_o]_i
 \, .
\label{eq:Vi}
\end{equation}
To each edge $e:i \to j$, we associate $(R_i, R_j)$ bi-module
\begin{equation}
 [\mathbf{H}_{e: i \to j}] =
% \frac{1}{ [\Lambda \bQ^{\vee}_{d_{ij}}]}[\bM_{e: i \to j}^{\vee}] [\bY_o]_i [\bY_o]_j^{\vee}
%  \frac{1}{ [\Lambda \bQ_{ij}^{\vee}] }
%  [\bM_{e: i \to j}^{\vee}]
  - \frac{1}{ [\Lambda \bQ_{ij}] }
  [\bM_{e: i \to j}]
  [\bY_o^\vee]_i [\bY_o]_j
\label{eq:Hij}
\end{equation}
where $d_{ij} %= \mathrm{min}(d_i,d_j)
= \mathrm{gcd}(d_i, d_j)$ and $\ch \Lambda \bQ_{ij} = (1-q_1^{d_{ij}})(1-q_2)$.
The character of $\bM_{e:i \to j}$ is given by the multiplicative mass parameter of the bifundamental hypermultiplet assigned to the edge $e: i \to j$ as $\ch \bM_{e:i \to j} = \mu_e$.
%Remark. The relation between $[\bX]_i$ and $[\bY_o]_i$ is the same as in simply-laced case. 
The observable $(\bY_o)_i$ is written in terms of $(\bX)_i$
\begin{align}
 [\bY_o]_i = [\Lambda \bQ_{1,i}] [\bX]_i
 \label{eq:Yo1}
\end{align}
where $[\bX]_i := [\bY_{\CalS_1}]_i$ is the $\CalS_2$-reduction of the space-time module $[\bY_{\CalS}]$ with $\CalS = \BC^2 = \CalS_1 \times \CalS_2$, and $\ch \Lambda \bQ_{1,i} = (1 - q_1^{d_i})$.
We can also apply another consistent path through the $\CalS_1$-reduction $[\tilde\bX]_i := [\bY_{\CalS_2}]_i$, which gives
\begin{align}
 [\bY_o]_i = [\Lambda \bQ_2] [\tilde\bX]_i
\end{align}
with $\ch \Lambda \bQ_2 = (1 - q_2)$ for $\forall i \in \Gamma_0$.
These two expressions are related through transposition of the partition $(\lambda_{i,\alpha})_{i \in \Gamma_0,\, \alpha \in [1 \ldots \sn_i]}$, labeling the $\sT$-fixed point.
Since $\CalS_1$ and $\CalS_2$ are not equivalent for a non-simply-laced quiver, this compatibility implies a nontrivial duality known as the quantum $q$-geometric Langlands duality~\cite{Frenkel:2010wd,Aganagic:2017smx}.

To describe the Chern character $X = \operatorname{ch}_\sT \bX$ at a $\sT$-fixed point, we introduce a set
\begin{align}
 \CalX_i & =
 \{ x_{i,\alpha,k} \}_{\alpha\in[1\ldots \sn_i], \, k\in[1\ldots\infty]}
 \, , \qquad
 x_{i,\alpha,k} = \nu_{i,\alpha} q_1^{d_i (k-1)} q_2^{\lambda_{i,\alpha,k}}
 \, , \qquad
 \CalX = \bigsqcup_{i \in \Gamma_0} \CalX_i
 \, .
\end{align}
We define
\begin{align}
 X_i = \sum_{x \in \CalX_i} x
 \, .
\end{align}
Thus a contribution to the Chern character of the observable sheaf from the node $i \in \Gamma_0$ is 
\begin{align}
 \ch \bY_i = (1 - q_1^{d_i}) X_i % \sum_{x \in \CalX_i} x
 \, ,
\label{eq:YX}
\end{align}
corresponding to \eqref{eq:Yo1}.  We denote the $p$-th Adams operation applied to $\bY_i$ by $\bY_i^{[p]}$.
The sheaves $(\bY_i^{[p]})_{i \in \Gamma_0, p \in \BZ_{\ge 1}}$ generate the ring of gauge theory observables.
The expression \eqref{eq:YX} implies the fractionalization
\begin{align}
 \ch \bY_i = (1 + q_1 + \cdots + q_1^{d_i-1}) \, \ch \by_i
 \label{eq:Y_factorization}
\end{align}
where the fractional observable sheaf is defined
\begin{align}
% \ch \by_i = (1 - q_1) X_i
 [\by]_i = [\Lambda \bQ_1] [X]_i
\end{align}
with $\ch \Lambda \bQ_1 = (1 - q_1)$.
This fractional sheaf plays a fundamental role in the geometric construction of fractionalization of Nakajima's quiver variety, which would be discussed in our forthcoming paper~\cite{Kimura:2018}.

The Chern characters of the vector and hypermultiplet contribution are now explicitly written as follows,
\begin{align}
 \ch \bV_i
 =
 \frac{1 - q_1^{-d_i}}{1 - q_2}
 \sum_{(x,x') \in \CalX_i^2} \frac{x'}{x}
 \, , \quad
 \ch \bH_{e:i \to j}
 =
 - \mu_e  %\frac{1 - q_1^{-1}}{1 - q_2^{d_{ij}}}
 \frac{(1 - q_1^{-d_i})(1 - q_1^{d_j})}{(1 - q_1^{d_{ij}})(1 - q_2)}
 \sum_{(x,x') \in \CalX_i \times \CalX_j} \frac{x'}{x}
 \, .
\end{align}
The total character is given in a compact form
\begin{align}
 \sum_{i \in \Gamma_0} \ch \bV_i + \sum_{e:i \to j} \ch \bH_{e:i \to j}
 & =
 \sum_{(x,x')\in\CalX^2}
 \left( c_{\si(x)\si(x')}^{+} \right)^\vee
 \frac{1 - q_1^{-d_{\si(x)}}}{1 - q_2}
 \frac{x'}{x}
 =
 \sum_{(x,x')\in\CalX^2}
 \left( b_{\si(x)\si(x')}^{+} \right)^\vee
 \frac{1 - q_1^{-1}}{1 - q_2}
 \frac{x'}{x} 
\end{align}
where $\si: \CalX \to \Gamma_0$ is the node label such that $\si(x) = i$ for $x \in \CalX_i$, and a half of the mass-deformed Cartan matrix is defined
\begin{align}
 [\bc_{ij}^+] & =
 \delta_{ij} - \sum_{e:i \to j} [\bM_e^\vee] \frac{[\Lambda \bQ_{1,i}^\vee]}{[\Lambda \bQ_{1,ij}^\vee]}
\end{align}
with $\ch \Lambda \bQ_{1,ij} = (1 - q_1^{d_{ij}})$, and its character is
\begin{align}
 c_{ij}^+ : = \ch \bc_{ij}^+ =
 & =
 \delta_{ij}
 - \sum_{e:i \to j} \mu_e^{-1} \frac{1 - q_1^{-d_j}}{1 - q_1^{-d_{ij}}}
 \nonumber \\ &
 =
 \delta_{ij}
 - \sum_{e:i \to j} \sum_{r = 0}^{d_j/d_{ij}-1} \mu_e^{-1} q_1^{-r d_{ij}}
 \ \stackrel{(c_{ij}^{+[0]})}{\longrightarrow} \
 \delta_{ij} - \#(e : i \to j) 
 \, ,
\label{eq:c_ijplus}
\end{align}
which coincides with a half of the ordinary Cartan matrix %\eqref{eq:Cartan_0th}
in the classical limit.
The number of edges is counted with the multiplicity $d_j/d_{ij}$,
\begin{align}
 \#(e : i \to j) = \sum_{e:i \to j} \frac{d_j}{d_{ij}}
 \label{eq:edge_num}
 \, .
\end{align}
Then the deformation of the (half of) symmetrized Cartan matrix is defined
\begin{align}
 [\bb_{ij}^+] = \frac{[\Lambda \bQ_{1,i}]}{[\Lambda \bQ_1]} [\bc_{ij}^+]
\end{align}
and its Chern character
%\eqref{eq:symCartan_0th}
\begin{align}
 b_{ij}^+ : = \ch \bb_{ij}^+
 = \frac{1 - q_1^{d_i}}{1 - q_1} c_{ij}^+
 = \frac{1 - q_1^{d_i}}{1 - q_1} \delta_{ij}
 - \sum_{e:i \to j} \mu_e^{-1} \frac{(1 - q_1^{d_i})(1 - q_1^{-d_j})}{(1 - q_1)(1 - q_1^{-d_{ij}})} 
 \, .
\end{align}
We also define $\left( c_{ij}^+ \right)^\vee := \ch \left( \bc_{ij}^+ \right)^\vee$, and $\left( b_{ij}^+ \right)^\vee := \ch \left( \bb_{ij}^+ \right)^\vee$.
If $d_i = 1$ for all $i \in \Gamma_0$ the definition of the deformed Cartan matrix agrees with
the one from~\cite{Kimura:2015rgi,Kimura:2016dys}. If the fractional quiver $(\Gamma,d)$ corresponds
to a non-simply-laced Lie algebra, our gauge theory definition of the $q_1$-dependent Cartan matrix corresponds to Frenkel--Reshetikhin's construction~\cite{Frenkel:1997} with $q_1=q_{\text{FR}}^2, q_2=t_{\text{FR}}^{-2}$. 

\subsection{Fractional quiver}
%\subsection{Non-simply-laced quiver}

%We use the notations of~\cite{Kimura:2015rgi,Kimura:2016dys}.
%
%Let $\Gamma$ be a quiver with the set of nodes $\Gamma_0$ and the set of arrows (edges) $\Gamma_1$.
%An edge from $i$ to $j$ is denoted by $e: i \to j$.

A quiver $\Gamma$ defines $|\Gamma_0| \times |\Gamma_0|$ matrix $(c_{ij})$, the mass-deformed Cartan matrix,
\begin{align}
 c_{ij}
 & =
 c_{ij}^+ + c_{ij}^-
 =
 ( 1 + q_{ii}^{-1}) \delta_{ij}
 - \sum_{e:i \to j} \mu_e^{-1} \frac{1 - q_1^{-d_j}}{1 - q_1^{-d_{ij}}}
 - \sum_{e:j \to i} \mu_e q_{ij}^{-1} \frac{1 - q_1^{-d_j}}{1 - q_1^{-d_{ij}}}
 \label{eq:c_ij}
\end{align}
where $(c_{ij}^+)$ is defined \eqref{eq:c_ijplus} and the other half matrix $(c_{ij}^-)$ is defined
\begin{align}
 c_{ij}^-
 & =
 q_{ii}^{-1} \delta_{ij}
 - \sum_{e:j \to i} \mu_e q_{ij}^{-1} \frac{1 - q_1^{-d_j}}{1 - q_1^{-d_{ij}}}
\end{align}
with $q_{ij} := q_1^{d_{ij}} q_2$ and  $q_{ii} = q_1^{d_i} q_2$.
In the classical limit, it is reduced to the quiver Cartan matrix
\begin{align}
 c_{ij} & = 2 \delta_{ij} - \#(e : i \to j) - \#(e : j \to i)
 \label{eq:Cartan_0th}
\end{align}
%which we call the quiver Cartan matrix.
where the number of edges $\#(e: i \to j)$ is meant with multiplicity $d_j/d_{ij}$ as in  \eqref{eq:edge_num}.
%\rem{If $(c_{ij})$ is symmetric, the quiver is simply-laced, and it has been studied in~\cite{Kimura:2015rgi,Kimura:2016dys}.
%Otherwise it is a non-simply-laced quiver.}
If there are no loops, all the diagonal elements are equal to 2, and such a matrix defines Kac--Moody algebra $\g(\Gamma)$ with Dynkin diagram $\Gamma$.

Similarly, symmetrization of the mass-deformed Cartan matrix \eqref{eq:c_ij} is defined
\begin{align}
 b_{ij}
 & = \frac{1 - q_1^{d_i}}{1 - q_1} c_{ij}
 \nonumber \\
 & = \frac{1 - q_1^{d_i}}{1 - q_1} (1 + q_{ii}^{-1}) \delta_{ij}
 - \sum_{e:i \to j} \mu_{e}^{-1}
 \frac{(1 - q_1^{d_i})(1 - q_1^{-d_j})}{(1 - q_1)(1 - q_1^{-d_{ij}})}
 - \sum_{e:j \to i} \mu_{e} q_{ij}^{-1}
 \frac{(1 - q_1^{d_i})(1 - q_1^{-d_j})}{(1 - q_1)(1 - q_1^{-d_{ij}})} 
 \, ,
 \label{eq:b_ij}
\end{align}
which obeys the reflection
\begin{align}
 b_{ij} = (q_1 q_2)^{-1} b_{ji}^\vee
 \, .
\end{align}
This definition agrees with the conventional definition of
the symmetrized Cartan matrix.

Let $c_{ij} = (\alpha_i^{\vee}, \alpha_j)$ be the symmetrizable Cartan
matrix where $(\alpha_j)$ is a system of simple roots, and $(\alpha_j^{\vee})$ is a system of simple
coroots, and let $(d_i)$ be positive integers such that the matrix
\begin{align}
 b_{ij} = d_i c_{ij}
 \label{eq:symCartan_0th}
\end{align}
is symmetric. We can choose a bilinear form on $\g$ such that 
\begin{align}
 d_i =(\alpha_i, \alpha_i)
 \, .
 \label{eq:root_length}
\end{align}
We remark that by Dynkin--Cartan ABCDEFG classification, for finite-dimensional Lie algebra $\g$, 
 if $c_{ij} \neq 0$, then $b_{ij} = \operatorname{max}(d_i, d_j)$. 

\subsection{Fractional quiver gauge theory partition function}

The vector and hypermultiplet contributions to the gauge theory partition function is  obtained as the index functor of the corresponding Chern character, which is the equivariant Witten index along a circle $S^1$ for 5d gauge theory on $\BR^4 \times S^1$.
In this paper we use the Dolbeault index
\begin{align}
 \BI\left[ \sum_k x_k \right]
 = \prod_k \left( 1 - x_k^{-1} \right)
 \label{eq:ind_func}
\end{align}
which obeys the reflection formula
\begin{align}
 \BI \left[ \bX^\vee \right]
 & =
 (-1)^{\rk \bX} \left( \det \bX \right)
 \BI \left[ \bX \right]
 \, .
 \label{eq:index_reflection}
\end{align}
%\begin{align}
% \exp
% \left(
%  \sum_{p=1}^\infty \frac{1}{p} (\bX^\vee)^{[p]}
% \right)
% & =
% (-1)^{\rk \bX} \left( \det \bX \right)
% \exp
% \left(
%  \sum_{p=1}^\infty \frac{1}{p} \bX^{[p]}
% \right)
% \label{eq:index_reflection}
%\end{align}
When the quiver gauge theory satisfies the conformal condition, the Dolbeault convention is equivalent to the Dirac index.
Otherwise we need a proper shift of Chern--Simons level.
The (full) partition functions are given by
\begin{align}
 Z_i^\text{vec}
 = \BI \left[ \bV_i \right]
 = \prod_{(x,x') \in \CalX_i^2}
 \left( q_1^{d_i} q_2 \frac{x}{x'}; q_2 \right)_\infty
 \left( q_2 \frac{x}{x'}; q_2 \right)_\infty^{-1}
 \, ,
\label{eq:Zvec}
\end{align}
and
\begin{align}
 Z_{e:i \to j}^\text{bf}
 & = \BI \left[ \bH_{e:i \to j} \right]
% = \prod_{(x,x') \in \CalX_i \times \CalX_j}
% \left( \mu_e^{-1} q_1 q_2^{d_{ij}} \frac{x}{x'}; q_2^{d_{ij}} \right)_\infty^{-1}
% \left( \mu_e^{-1} q_2^{d_{ij}} \frac{x}{x'}; q_2^{d_{ij}} \right)_\infty
% \nonumber \\
 & =
 \prod_{(x,x') \in \CalX_i \times \CalX_j}
 \prod_{r=0}^{d_j/d_{ij}-1}
 \left( \mu_e^{-1} q_1^{-r d_{ij}} q_1^{d_{i}} q_2 \frac{x}{x'}; q_2 \right)_\infty^{-1}
 \left( \mu_e^{-1} q_1^{-r d_{ij}} q_2 \frac{x}{x'}; q_2 \right)_\infty 
 \, .
\label{eq:Zhyp}
\end{align}
In particular, the bifundamental factor exhibits a peculiar behavior depending on $(d_i)_{i \in \Gamma_0}$: There appear the additional contributions with the duplicated mass parameters $(\mu_{e:i \to j} q_1^{r d_{ij}})$ for $r \in [0 \ldots d_j/d_{ij}-1]$, which is similar to that found in 3d non-simply-laced quiver gauge theory~\cite{Dey:2016qqp}.
Replacing the index \eqref{eq:ind_func} with the equivariant elliptic genus with respect to two-torus $T^2$ with modulus $\tau$
\begin{align}
 \BI_p \left[ \sum_k x_k \right]
 & =
 \prod_{k} \theta(x^{-1}_k;p)
\end{align}
where  $p = \exp \left( 2 \pi \iota \tau\right)$ is multiplicative modulus and 
\begin{align}
 \theta(x;p) = (x;p)_\infty (px^{-1};p)_\infty
 \, ,
\end{align}
we obtain the 6d gauge theory partition function on $\BR^4 \times T^2$, which yields the elliptic deformation of W-algebra~\cite{Kimura:2016dys}.
We remark that the elliptic index obeys the same reflection formula \eqref{eq:index_reflection} as well, and the conformal condition is mandatory for 6d theory to avoid the modular/gauge anomaly.

Then we introduce conjugate variables to the local observables $(t_{i,p})_{i \in \Gamma_0, p \in \BZ_{\ge 1}}$, called the higher time variables like in the integrable hierarchy~\cite{Marshakov:2006ii}, so that the partition function plays a role of the generating function of the observables $(\bY_i^{[p]})_{i \in \Gamma_0, p \in \BZ_{\ge 1}}$.
See also~\cite{Nakajima:2003uh}.
Together with the Chern--Simons levels assigned to each node $(\kappa_i)_{i \in \Gamma_0}$, the gauge theory partition function is obtained as the summation over the $\sT$-fixed point of the moduli space
%\rem{$b$-matrix used}
\begin{align}
 Z_\sT(t)
 & =
 \sum_{\CalX \in \frakM^\sT}
 \exp
 \left(
 \sum_{(x,x')\in\CalX^2}
 \sum_{p=1}^\infty - \frac{1}{p}
 \frac{1 - q_1^{p}}{1 - q_2^{-p}}
 \left( b_{\si(x)\si(x')}^{+[p]}\right)
 \frac{x^p}{x'^p}
 \right)
 \nonumber \\
 & \quad \times
 \exp
 \left(
 \sum_{x \in \CalX}
 \left(
 - \frac{\kappa_{\si(x)}}{2}
 \log_{q_2} x \left( \log_{q_2} x - 1 \right)
 + \log \fq_{\si(x)} \log_{q_2} \frac{x}{\mathring{x}}
 + \sum_{p=1}^\infty (1 - q_1^{d_{\si(x)}p}) \, t_{\si(x),p} \, x^p
 \right)
 \right)
 \, .
 \label{eq:Z-function}
\end{align}
Here $\fq_i$ is the gauge coupling for the node $i$.
The instanton number, which counts the size of partition $\lambda$, is given by
\begin{align}
 \sk_i %= |\lambda_i|
 = 
 \sum_{x \in \CalX_i, \mathring{x} \in \CalX_{0,i}}
 \log_{q_2} \frac{x}{\mathring{x}}
\end{align}
where the ground configuration, corresponding to empty partition $\lambda = \emptyset$, is defined
\begin{align}
 \mathring{x}_{i,\alpha,k} = \nu_{i,\alpha} q_1^{d_i(k-1)}
 \, .
\end{align}
$\displaystyle
\CalX_{0,i} = \{ \mathring{x}_{i,\alpha,k} \}_{i\in\Gamma_0,\, \alpha\in[1\ldots \sn_i], \, k\in[1\ldots\infty]}$ is a set of such ground configuration, and $\CalX_0 = \bigsqcup_{i \in \Gamma_0} \CalX_{0,i}$.
The 6d theory partition function has a similar expression.
See \cite{Kimura:2016dys} for details.

\section{Operator formalism}

\subsection{$Z$-state}

Since the $t$-extended partition function \eqref{eq:Z-function} plays a role of the generating function, the (non-normalized) average of the gauge theory observable is given by
\begin{align}
 \Big< \bY_i^{[p]} \Big>
 & =
 \frac{\partial}{\partial t_{i,p}} Z_\sT(t)\Bigg|_{t=0}
 \, .
\end{align}
From this point of view, the observable is equivalent to the derivative with the time variable, and thus identified as an operator obeying the Heisenberg algebra,
\begin{align}
 \left[ \frac{\partial}{\partial t_{i,p}}, t_{j,p'} \right]
 = \delta_{ij} \delta_{pp'}
 \, .
\end{align}
The $t$-extended partition function, which explicitly depends on the operators $(t_{i,p})_{i \in \Gamma_0, p \in \BZ_{\ge 1}}$, can be treated as an operator in the free field formalism. 
To this operator we can associate a state in the Fock space generated by action of the Heisenberg algebra on the vacuum, like in the operator-state correspondence in conformal field theory.

We define the $Z$-state using the screening current operator
\begin{align}
 \ket{Z_\sT}
 & =
 \sum_{\CalX \in \frakM^\sT}
 \prod_{x \in \CalX}^\succ
 S_{\si(x),x} \ket{1}
 \label{eq:Z-state}
\end{align}
where the product is radial-ordered with respect to the parameter $x \in \BC^{\times}$.
The vacuum state $\ket{1}$ is annihilated by all the derivative operators $(\partial/\partial t_{i,p})_{i \in \Gamma_0, p \in \BZ_{\ge 1}}$, and the screening current is defined
\begin{align}
 S_{i,x}
 & = \
 :\exp \left(
 s_{i,0} \log x + \tilde{s}_{i,0}
% + \frac{\kappa_i}{2} \log_{q_2^{d_i}} x
% \left( \log_{q_2^{d_i}} x - 1 \right)
 + \sum_{p \neq 0} s_{i,p} x^{-p}
 \right):
\end{align}
where the free field oscillators are 
\begin{align}
 s_{i,-p}
 \stackrel{p > 0}{=}
 (1 - q_1^{d_i p}) t_{i,p}
 \, , \quad
 s_{i,0} = t_{i,0}
 \, , \quad
 \tilde{s}_{i,0} = -\beta c_{ji}^{[0]} \frac{\partial}{\partial t_{j,0}}
 \, , \quad 
 s_{i,p}
 \stackrel{p > 0}{=}
 - \frac{1}{p} \frac{1}{1 - q_2^{-p}} c_{ji}^{[p]}
 \frac{\partial}{\partial t_{j,p}}
 \, ,
\end{align}
with the commutation relation
\begin{align}
 \Big[ s_{i,p}, s_{j,p'} \Big]
 = - \frac{1}{p} \frac{1 - q_1^{d_j p}}{1 - q_2^{-p}} \, c_{ji}^{[p]} \,
 \delta_{p+p',0}
 = - \frac{1}{p} \frac{1 - q_1^{p}}{1 - q_2^{-p}} \, b_{ji}^{[p]} \, \delta_{p+p',0}
 \, ,
\end{align} 
\begin{align}
 \Big[ \tilde{s}_{i,0}, s_{j,p} \Big]
 = - \beta \, c_{ji}^{[0]} \, \delta_{p,0}
 \, , \qquad
 \beta = - \frac{\log q_1}{\log q_2}
 \, .
\end{align}
The matrices $(c_{ij}^{[p]})$ and $(b_{ij}^{[p]})$ are obtained from the $p$-th Adams operation of the mass-deformed total Cartan matrix~\eqref{eq:c_ij} and its symmetrization \eqref{eq:b_ij}.

The $Z$-state in the operator formalism \eqref{eq:Z-state} is computed using the free field operators
\begin{align}
 Z_\sT(t)
 & =
 \sum_{\CalX \in \frakM^\sT}
 \exp
 \left(
 \sum_{(x \succ x') \in \CalX^2} \sum_{p=1}^\infty
 - \frac{1}{p}
 \frac{1 - q_1^{p}}{1 - q_2^{-p}}
 \left( b_{\si(x)\si(x')}^{[p]}\right) \frac{x^p}{x'^p}
 \right)
 \nonumber \\
 & \quad \times
 \exp
 \left(
 \sum_{x \in \CalX}
 \left(
 - \frac{\kappa_{\si(x)}}{2}
 \log_{q_2} x \left( \log_{q_2} x - 1 \right)
 + \log \fq_{\si(x)} \log_{q_2} \frac{x}{\mathring{x}}
 + \sum_{p=1}^\infty (1 - q_1^{d_{\si(x)} p}) \, t_{\si(x),p} \, x^p
 \right)
 \right) 
\end{align}
which is obtained as a summation over the pair contributions under the ordering $(x \succ x')$.
Due to the reflection formula \eqref{eq:index_reflection}, it coincides with the gauge theory definition of the partition function \eqref{eq:Z-function} evaluated as
\begin{align}
 \kappa_i = - \sn_j (c_{ji}^-)^{[0]}
 \, , \qquad
 \log_{q_2} \fq_i
 =
 \beta + t_{i,0} + \sn_j (c_{ji}^-)^{[\log_{q_2}]}
 - \log_{q_2} ( (-1)^{\sn_j} \nu_j) (c_{ji}^-)^{[0]} 
\end{align}
where
\begin{align}
 (c_{ji}^-)^{[\log_{q_2}]}
 & =
 \delta_{ij} \log_{q_2} q_{ii}^{-1}
 - \sum_{e:i \to j}
 \log_{q_2}
 \left(
 \mu_e q_{ij}^{-1} \frac{1 - q_2^{-d_i}}{1 - q_2^{-d_{ij}}}
 \right)
 \, .
\end{align}

\subsection{Screening charge}

The gauge theory partition function is given as an infinite sum over the moduli space fixed point $\frakM^\sT$.
The summation in the $Z$-state \eqref{eq:Z-state} is replaced with that over $\BZ^{\CalX_0}$, which is a set of arbitrary integer sequences terminating by zeros (see~\cite{Kimura:2015rgi}):
\begin{align}
 \ket{Z_\sT} & =
 \sum_{\CalX \in \BZ^{\CalX_0}}
 \prod_{x \in \CalX}^\succ
 S_{\si(x),x} \ket{1}
 \, ,
 \label{eq:Z-state2}
\end{align}
because there appears a zero factor for $\CalX \in \BZ^{\CalX_0}$, but $\CalX\not\in \frakM^\sT$,
\begin{align}
 \prod_{x \in \CalX}^\succ S_{\si(x),x} \ket{1} = 0
 \, .
\end{align}
Introducing the screening charge operator
\begin{align}
 \sS_{i,\mathring{x}}
 & =
 \sum_{s_2 \in \BZ} S_{i,q_2^{s_2} \mathring{x}}
 \, ,
\end{align}
the $Z$-state is obtained as an ordered product
\begin{align}
 \ket{Z_\sT}
 & =
 \prod_{\mathring{x} \in \CalX_0}^\succ \sS_{\si(\mathring{x}),\mathring{x}}
 \ket{1}
 \, .
\end{align}
The vacuum $\ket{1}$ of the Heisenberg algebra $\bH$ is a constant with respect to the time variables $(t_{i,p})$, obeying $(\partial/\partial t_{i,p})\ket{1} = 0$ for $i \in \Gamma_0, p \in \BZ_{\ge 1}$.
Its dual $\bra{1}$ plays a role of the projector to the $t=0$ sector because $\bra{1}t_{i,p} = 0$ for $i \in \Gamma_0, p \in \BZ_{\ge 1}$.
Thus the non-$t$-extended (plain) partition function is given as a correlator of the screening charges (see also~\cite{Aganagic:2013tta,Aganagic:2014oia,Aganagic:2015cta})
\begin{align}
 Z(t=0) = \vev{1|Z_\sT}
 & =
 \bra{1}
 \prod_{\mathring{x} \in \CalX_0}^\succ \sS_{\si(\mathring{x}),\mathring{x}}
 \ket{1}
 \label{eq:Z-correlator}
\end{align}

\subsection{$\sV$-operator: fundamental matter}\label{sec:fund_mat}

In addition to the vector and bifundamental hypermultiplet, we can also consider the (anti)fundamental hypermultiplet.
It is obtained from the bifundamental matter connecting with the flavor node, whose gauge coupling is turned off.
Such an additional contribution can be reproduced by the $\sV$-operator acting on the gauge theory $Z$-state.

We define the $\sV$-operator 
\begin{align}
 \sV_{i,x} & =
 \exp \left( \sum_{p \neq 0} v_{i,p} \, x^{-p} \right)
\end{align}
where the free field operator defined
\begin{align}
 v_{i,-p}
 \stackrel{p>0}{=}
 - \tilde{c}_{ij}^{[-p]} t_{j,p} 
 \, , \qquad
 v_{i,p}
 \stackrel{p>0}{=}
 \frac{1}{p} \frac{1}{(1 - q_1^{d_i p})(1 - q_2^{p})}
 \frac{\partial}{\partial t_{i,p}}
 \, .
\end{align}
Thus the $\sV$-operator $\sV_{i,\mu}$ generates the shift of the time variables
\begin{align}
 t_{i,p}
 \ \longrightarrow \
 t_{i,p} + \frac{1}{p} \frac{1}{(1 - q_1^{d_i p})(1 - q_2^{p})} \mu^{-p}
 \, .
\end{align}
The commutation relation between $v$ and $s$ oscillators is given by
\begin{align}
 \Big[
  v_{i,p}, s_{j,p'}
 \Big]
 =
 \frac{1}{p} \frac{1}{1 - q_2^{p}} \delta_{ij} \delta_{p+p',0}
\end{align}
which yields the OPE with the screening current 
\begin{align}
 \sV_{i,x} S_{i,x'}
 =
 \left( \frac{x'}{x}; q_2 \right)_\infty^{-1}
 : \sV_{i,x} S_{i,x'} :
 \, , \qquad
 S_{i,x'} \sV_{i,x}
 =
 \left( q_2 \frac{x}{x'}; q_2 \right)_\infty
 : \sV_{i,x} S_{i,x'} :
 \, .
\end{align}
These OPE factors provide the fundamental and anti-fundamental hypermultiplet contributions.
The $t$-extended $Z$-state in the presence of these matter contributions is given by
\begin{align}
 \ket{Z_\sT}
 & =
 \left( \prod_{x \in \CalX_\text{f}} \sV_{\si(x),x} \right)
 \left(
 \prod_{\mathring{x} \in \CalX}^\succ \sS_{\si(\mathring{x}),\mathring{x}}
 \right)
 \left( \prod_{x \in \tilde\CalX_\text{f}} \sV_{\si(x),x} \right)
 \ket{1}
\end{align}
where $\CalX_\text{f} = \{\mu_{i,f}\}_{i \in \Gamma_0,f\in[1\ldots \sn_i^\text{f}]}$ and $\tilde\CalX_\text{f} = \{\tilde\mu_{i,f}\}_{i \in \Gamma_0,f\in[1\ldots \tilde\sn_i^\text{f}]}$ are sets of the multiplicative fundamental and antifundamental mass parameters.
The $\sV$-operator creates a pole singularity on the curve at $x = \mu_{i,f}$, which is consistent with the Seiberg--Witten geometry perspective.
Then the non-extended partition function is given as a correlator with additional $\sV$-operators inserted,
\begin{align}
 Z_\sT(t=0)
 & =
 \bra{1}
 \left( \prod_{x \in \CalX_\text{f}} \sV_{\si(x),x} \right)
 \left(
 \prod_{\mathring{x} \in \CalX}^\succ \sS_{\si(\mathring{x}),\mathring{x}}
 \right)
 \left( \prod_{x \in \tilde\CalX_\text{f}} \sV_{\si(x),x} \right)
 \ket{1}
 \, .
\end{align}

\subsection{$\sY$-operator: generating current of observables}

In addition to the screening current operator used to construct the $Z$-state, we define another operator, called the $\sY$-operator,
\begin{align}
 \sY_{i,x}
 & =
 q_1^{d_i \tilde{\rho}_i}
 :\exp
 \left(
 y_{i,0} + \sum_{p \neq 0} y_{i,p} \, x^{-p}
 \right)
 :
\end{align}
with the Weyl vector $\displaystyle \tilde{\rho}_i = \sum_{j \in \Gamma_0} \tilde{c}_{ji}^{[0]}$, and $\left(\tilde{c}_{ij}\right)$ is the inverse of the Cartan matrix if it is invertible.
If it is not invertible, we have to deal with the $q_1$ factor separately.
The free field oscillators are defined
\begin{align}
\label{eq:yip}
 y_{i,-p}
 \stackrel{p>0}{=}
 (1 - q_1^{d_i p}) (1 - q_2^{p}) \tilde{c}_{ji}^{[-p]} t_{j,p}
 \, , \qquad
 y_{i,0} = - \tilde{c}_{ji}^{[0]} t_{j,0} \log q_2%^{d_i}
 \, , \qquad
 y_{i,p}
 \stackrel{p>0}{=}
 - \frac{1}{p} \frac{\partial}{\partial t_{i,p}}
\end{align}
obeying the commutation relation
\begin{align}
 \Big[ y_{i,p}, y_{j,p'} \Big]
 =
 - \frac{1}{p} (1 - q_1^{d_j p}) (1 - q_2^p)
 \, \tilde{c}_{ij}^{[-p]} \, \delta_{p+p',0}
 \, .
\end{align}
The commutation relation for $(y_{i,p})_{i \in \Gamma_0}$ and $(s_{j,p'})_{j \in \Gamma_0}$ is then given by
\begin{align}
 \Big[
  y_{i,p}, s_{j,p'}
 \Big]
 =
 - \frac{1}{p} (1 - q_1^{d_i p}) \, \delta_{ij} \, \delta_{p+p',0}
 \, , \qquad
 \Big[
  \tilde{s}_{i,0}, y_{j,0}
 \Big]
 = - \delta_{ij} d_i \log q_1
 \, ,
\end{align}
which leads to the ordered product
\begin{align}
 |x| > |x'| :
 & \quad
 \sY_{i,x} S_{j,x'}
 = \ : \sY_{i,x} S_{j,x'} :
 \begin{cases}
  \displaystyle
  \frac{1 - x'/x}{1 - q_1^{d_i} x'/x} & (i = j) \\[1em]
  1 & (i \neq j)
 \end{cases}
% \, , \qquad
% : \sY_{i,x} S_{j,x'} : \quad (i \neq j)
 \, , \\
 |x| < |x'| :
 & \quad
 S_{j,x'} \sY_{i,x} 
 = \ : \sY_{i,x} S_{j,x'} :
 \begin{cases}
  \displaystyle
  q_1^{-d_i} \frac{1 - x/x'}{1 - q_1^{-d_i} x/x'}  & (i = j) \\[1em]
  1 & (i \neq j)
 \end{cases}
 \, . 
\end{align}
There is a pole at $x = q_1^{d_i} x'$ in the product for $i = j$, and thus the commutation relation between the $\sY$-operator and the screening current is given by
\begin{align}
 \Big[
  \sY_{i,x}, S_{j,x'}
 \Big]
 =
 \begin{cases}
  \displaystyle
  (1 - q_1^{-d_i}) \, \delta \left( q_1^{d_i} \frac{x'}{x} \right)
  : \sY_{i,x}, S_{j,x'} : & (i = j) \\[1em]
  0 & (i \neq j)
 \end{cases}
\end{align}
where the delta function is defined
\begin{align}
 \delta(x)
 & =
 \sum_{p \in \BZ} x^p
 % = \frac{1}{1-x} + \frac{x^{-1}}{1 - x^{-1}}
 \, .
\end{align}
Thus the $\sY$-operator commutes with the screening current in the limit $q_1 \to 1$.
The $\sY$-operator average in the non-$t$-extended gauge theory is represented as a correlator as well as the partition function \eqref{eq:Z-correlator},
\begin{align}
 \bra{1} \sY_{i,x} \prod_{x' \in \CalX}^\succ S_{\si(x),x'} \ket{1}
 & =
 q_1^{d_i \tilde{\rho}_i}
 \left( \prod_{x' \in \CalX_i} \frac{1 - x'/x}{1 - q_1^{d_i} x'/x} \right)
 \bra{1} \prod_{x' \in \CalX}^\succ S_{\si(x),x'} \ket{1}
 \, .
\end{align}
Since the infinite product is written as
\begin{align}
 \prod_{x' \in \CalX_i} \frac{1 - x'/x}{1 - q_1^{d_i} x'/x}
 & =
 \exp \left(
  \sum_{p=1}^\infty - \frac{x^{-p}}{p} \bY_{i}^{[p]}
 \right)
 \, ,
\end{align}
the $\sY$-operator is the generating current of the gauge theory observable $(\bY_{i}^{[p]})_{i \in \Gamma_0, p \in \BZ_{\ge 1}}$, which is consistent with the definition given in~\cite{Nekrasov:2013xda}.
In addition, it is also possible to write in terms of the fractional observables, due to the factorization \eqref{eq:Y_factorization},
\begin{align}
 \exp \left(
  \sum_{p=1}^\infty - \frac{x^{-p}}{p} \bY_{i}^{[p]}
 \right)
 =
 \prod_{r=0}^{d_i-1}
  \exp \left(
  \sum_{p=1}^\infty - \frac{(q_1^{-r}x)^{-p}}{p} \by_{i}^{[p]}
 \right)
 \, .
\end{align}

\subsection{$\sA$-operator: iWeyl reflection}

Since the screening charge is given as a summation over the screening current, it is explicitly invariant under the $\BZ$-shift, $s_2 \to s_2 + \BZ$.
Correspondingly the gauge theory partition function has the corresponding $\BZ$-shift symmetry, which is also interpreted as change of variables.
To see the behavior of the partition function under the $\BZ$-shift, we define the $\sA$-operator 
\begin{align}
 \sA_{i,x} = \
 q_1^{d_i} : \frac{S_{i,x}}{S_{i,q_2 x}} :
 \, .
\end{align}
The free field representation is given by
\begin{align}
 \sA_{i,x}
 & =
 q_1^{d_i}
 :
 \exp
 \left( a_{i,0} + \sum_{p \neq 0} a_{i,p} \, x^{-p} \right)
 :
\end{align}
where the oscillators are defined
\begin{align}
 a_{i,p} = (1 - q_2^{-p}) s_{i,p}
 \, , \qquad
 a_{i,0} = - t_{i,0} \log q_2
 \, .
\end{align}
Since the $a$-oscillator is related to the $y$-oscillator using the Cartan matrix,
\begin{align}
 a_{i,p} = y_{j,p} \, c_{ji}^{[p]}
\end{align}
the $\sA$-operator plays a role as ``root'', while the $\sY$-operator is ``weight'', which is written in terms of the $\sY$-operators,
\begin{align}
 \sA_{i,x} & = \
 :
 \sY_{i,x} \sY_{i,q_{ii}x}
 \left(
 \prod_{e:i \to j} \prod_{r=0}^{d_j/d_{ij}-1}
 \sY_{j,\mu_e q_1^{r d_{ij}}x}
 \prod_{e:j \to i} \prod_{r=0}^{d_j/d_{ij}-1}
 \sY_{j,\mu_e^{-1} q_{ij} q_1^{r d_{ij}}x}   
 \right)^{-1}
 :
 \, .
\end{align}

\subsubsection{$qq$-character generated by the reflection}

The pole singularity of the $\sY$ \& $S$ product is canceled in the following combination,
\begin{align}
 \Res{x' \to q_1^{-d_i} x}
 \left[
 \sY_{i,x} S_{i,x'} \
 + :\sY_{i,x} \sA^{-1}_{i,q_{ii}^{-1} x}: S_{i,q_2^{-1}x}
 \right]
 = 0
 \, .
\end{align}
Here the $\sA$-operator plays a role of the generator of the iWeyl reflection~\cite{Nekrasov:2015wsu}.
In terms of the $\sY$-operators, the reflection is given by
\begin{align}
 \sY_{i,q_{ii} x}
 \ \longrightarrow \
 : \sY_{i,q_{ii} x} \sA_{i,x}^{-1} :
 \ = \ :
 \sY_{i,x}^{-1}
 \prod_{e:i \to j} \prod_{r=0}^{d_j/d_{ij}-1}
 \sY_{j,\mu_e q_1^{r d_{ij}}x}
 \prod_{e:j \to i} \prod_{r=0}^{d_j/d_{ij}-1}
 \sY_{j,\mu_e^{-1} q_{ij} q_1^{r d_{ij}}x}
 :
 \, .
\end{align}

Therefore the $qq$-character generated by the iWeyl reflection
\begin{align}
 T_{i,x} = \sY_{i,x} \ + \ : \sY_{i,x} \sA_{i,q_{ii}^{-1} x}^{-1} : + \cdots
\end{align}
does not have any pole singularities, and commutes with the screening charge
\begin{align}
 \Big[ T_{i,x}, \sS_{j,x'} \Big] = 0
 \, .
\end{align}
This assures the regularity of the $Z$-state of $t$-extended gauge theory, and holomorphy of the $qq$-character,
\begin{align}
 \partial_{\bar{x}} T_{i,x} \ket{Z_\sT} = 0
 \, .
\end{align}

\subsubsection{Collision and derivative term}

If there is a product of the $\sY$-operators which belong to the same node $i \in \Gamma_0$, we need an extra factor,
\begin{align}
 :\sY_{i,x} \sY_{i,x'}:
 + \
 \msS_{d_i} \left( \frac{x'}{x} \right)
 : \frac{\sY_{i,x} \sY_{i,x'}}{\sA_{i,q_{ii}^{-1} x}} :
 + \
 \msS_{d_i} \left( \frac{x}{x'} \right)
 : \frac{\sY_{i,x} \sY_{i,x'}}{\sA_{i,q_{ii}^{-1} x'}} :
 +
 : \frac{\sY_{i,x} \sY_{i,x'}}
        {\sA_{i,q_{ii}^{-1} x}\sA_{i,q_{ii}^{-1} x'}} :
\end{align}
where 
\begin{align}
 \msS_k(x)
 & =
 \frac{(1 - q_1^k x)(1 - q_2 x)}{(1 - x)(1 - q_1^k q_2 x)}
 =
 \exp
 \left(
  \sum_{p=1}^\infty \frac{1}{p} (1 - q_1^{k p})(1 - q_2^{p}) x^p
 \right)
 \, ,
 \label{eq:S-factor}
\end{align}
which corresponds to the OPE of $\sY$ and $\sA$ operators.
In particular, we write $\msS(x)=\msS_1(x)$ for simplicity, and remark the formula
\begin{align}
 \msS_k(x) = \prod_{r=0}^{k-1} \msS(q_1^r x)
 \, .
\end{align}

In the limit $x' \to x$, we have a derivative term
\begin{align}
 :\sY_{i,x}^2:
 + \
 :
 \left(
 \mathfrak{c}_i(q_1,q_2) - \frac{(1 - q_1^{d_i})(1 - q_2)}{1 - q_{ii}}
 \partial_{\log x} \log \sA_{i,q_{ii}^{-1} x}
 \right)
 \frac{\sY_{i,x}^2}{\sA_{i,q_{ii}^{-1} x}} :
 +
 : \frac{\sY_{i,x}^2}
 {\sA_{i,q_{ii}^{-1} x}^2} :
 \, ,
\end{align}
and the constant is defined
\begin{align}
 \mathfrak{c}_i(q_1,q_2)
 = \lim_{x \to 1} \left( \msS_{d_i}(x) + \msS_{d_i}(x^{-1}) \right)
 \, .
\end{align}
We remark, in the Nekrasov--Shatashvili limit $q_{1,2} \to 1$, the derivative term vanishes, due to the factor $(1 - q_1^{d_i})(1 - q_2)$.
We can similarly consider the higher-degree collision term $:\sY_{i,x}^n:$, which correspondingly involves higher derivatives of the $\sA$-operator.

\section{Fractional quiver W-algebras}

As shown in the previous section, we have a regular holomorphic current in the $t$-extended quiver gauge theory
\begin{align}
 \partial_{\bar{x}} T_{i,x} \ket{Z_\sT} = 0
\end{align}
where the operator $T_{i,x}$ is given as the $qq$-character generated by the iWeyl reflection.
The regularity of the current is equivalent to the commutation relation with the screening charge
\begin{align}
 \Big[ T_{i,x}, \sS_{j,x'} \Big] = 0
 \qquad
 \forall j \in \Gamma_0
 \, .
\end{align}
Thus the operator $T_{i,x}$ is a well-defined conserved current with the time-independent modes
\begin{align}
 T_{i,x} = \sum_{p \in \BZ} T_{i,p} \, x^{-p}
 \, .
\end{align}
The algebra generated by the holomorphic current $T_{i,x}$ defines the W($\Gamma$)-algebra associated with quiver $\Gamma$, which is constructed with the free field operators from the Heisenberg algebra $\bH$.
The $qq$-character defines the holomorphic generating current of W($\Gamma$)-algebra in the free field representation.

\subsection{$BC_2$ quiver}\label{sec:BC2}

The simplest example is $BC_2$ quiver:
\begin{center}
 \begin{tikzpicture}[thick,baseline=0]

  \draw (0,.15) circle [radius = .67em] node (1) {$2$};
  
  \draw (1,.15) circle [radius = .67em] node (2) {$1$};
  
  \draw (1) -- (2);

  \node at (0,-.5) {$1$};
  \node at (1,-.5) {$2$};

  \node at (-.5,-.45) [left] {node:};
  
 \end{tikzpicture}
\end{center}
where the integers assigned to each node is the root length \eqref{eq:root_length}, namely $d_1 = 2,
d_2 = 1$. This is different from the standard notation
\begin{tikzpicture}[thick,baseline=0]

 \draw [double,double distance = 2pt] (0,.15) -- ++(1,0); %(1) -- (2);
 
 \filldraw[fill = white] (0,.15) circle [radius = .35em] node (1) {};

 \filldraw[fill = white] (1,.15) circle [radius = .35em] node (2) {};

 \draw (.4,.35) -- (.6,.15) -- (.4,-.05);
 
\end{tikzpicture} for $BC_2$ quiver.

The mass-deformed Cartan matrix is
\begin{align}
 (c_{ij})
 & =
 \begin{pmatrix}
  1 + q_1^{-2} q_2^{-1} & - \mu^{-1} \\
  - \mu q_1^{-1} q_2^{-1} (1 + q_1^{-1}) & 1 + q_1^{-1} q_2^{-1}
 \end{pmatrix}
 \ \stackrel{(c_{ij}^{[0]})}{\longrightarrow} \
 \begin{pmatrix}
  2 & -1 \\ -2 & 2
 \end{pmatrix}
\end{align}
where the multiplicative bifundamental mass parameter is defined
\begin{align}
 \mu := \mu_{1 \to 2} = \mu_{2 \to 1}^{-1} q_1 q_2
 \, .
 \label{eq:BC_mass}
\end{align}
The $qq$-characters are generated by the local iWeyl reflection
\begin{align}
 \sY_{1,x}
 \ \longrightarrow \
 \frac{\sY_{2,\mu^{-1}x} \sY_{2,\mu^{-1}q_1^{-1}x}}
      {\sY_{1,q_1^{-2}q_2^{-1}x}}
 \, , \qquad
 \sY_{2,x}
 \ \longrightarrow \
 \frac{\sY_{1,\mu q_1^{-1} q_2^{-1} x}}{\sY_{2,q_1^{-1}q_2^{-1}x}}
\end{align}
which yields
\begin{align}
 T_{1,x}
 & =
 \sY_{1,x}
 + \frac{\sY_{2,\mu^{-1}x} \sY_{2,\mu^{-1}q_1^{-1}x}}
        {\sY_{1,q_1^{-2} q_2^{-1} x}}
 + \msS(q_1) \frac{\sY_{2,\mu^{-1}x}}{\sY_{2,\mu^{-1}q_1^{-2}q_2^{-1}x}}
 + \frac{\sY_{1,q_1^{-1} q_2^{-1} x}}
        {\sY_{2,\mu^{-1} q_1^{-1} q_2^{-1} x} \sY_{2,\mu^{-1} q_1^{-2} q_2^{-1} x}}
 + \frac{1}{\sY_{1,q_1^{-3} q_2^{-2} x}}
 \, , \\
 T_{2,x}
 & =
 \sY_{2,x}
 + \frac{\sY_{1,\mu q_1^{-1} q_2^{-1} x}}{\sY_{2,q_1^{-1} q_2^{-1} x}}
 + \frac{\sY_{2,q_1^{-2} q_2^{-1} x}}{\sY_{1,\mu q_1^{-3} q_2^{-2} x}}
 + \frac{1}{\sY_{2,q_1^{-3} q_2^{-2} x}}
 \, ,
\end{align}
where
\begin{align}
 \msS(q_1) & =
 \frac{(1 + q_1)(1 - q_1 q_2)}{1 - q_1^2 q_2}
 \, .
\end{align}
These characters correspond to the \textbf{5} (vector) and \textbf{4} (spinor) representations.
Here we omit the normal ordering symbol as long as no confusion.
We remark that the $\msS$-factor \eqref{eq:S-factor} appears in the first current $T_{1,x}$ at the zero weight term.
These holomorphic currents obey the OPE
\begin{align}
 &
 f_{11}\left(\frac{y}{x}\right) T_{1,x} T_{1,y}
 - f_{11}\left(\frac{x}{y}\right) T_{1,y} T_{1,x}
 \nonumber \\
 & % \hspace{8em}
 =
 - \frac{(1 - q_1^2)(1 - q_2)}{1 - q_1^2 q_2}
 \left(
 \delta \left( q_1^2 q_2 \frac{y}{x} \right)
 f_{22} \left( q_1^{-1} \right)
 T_{2,\mu^{-1} x} T_{2,\mu^{-1} q_1^{-1} x}
 \right.
 \nonumber \\
 & \hspace{14em}
 \left.
 - \delta \left( q_1^{-2} q_2^{-1} \frac{y}{x} \right)
 f_{22} \left( q_1 \right)
 T_{2,\mu^{-1} q_1 q_2 x} T_{2,\mu^{-1} q_1^2 q_2 x}
 \right)
 \nonumber \\[.5em]
 & \quad
 - \frac{(1 - q_1^2)(1 - q_2)(1 - q_1 q_2^2)(1 - q_1^3 q_2)}
        {(1 - q_1 q_2)(1 - q_1^2 q_2)(1 - q_1^3 q_2^2)} 
 % \msS_1 (q_1 q_2)
 \left(
 \delta \left( q_1^3 q_2^2 \frac{y}{x} \right)
 - \delta \left( q_1^{-3} q_2^{-2} \frac{y}{x} \right)
 \right)
\end{align}
\begin{align}
 &
 f_{12} \left( \frac{y}{x} \right) T_{1,x} T_{2,y}
 - f_{21} \left( \frac{x}{y} \right) T_{2,y} T_{1,x}
 \nonumber \\
 & \hspace{4em}
 =
 - \frac{(1 - q_1^2)(1 - q_2)}{1 - q_1^2 q_2}
 \left(
 \delta \left(\mu q_1^2 q_2 \frac{y}{x} \right) T_{2,\mu^{-1} x}
 - \delta \left(\mu q_1^{-3} q_2^{-2} \frac{y}{x} \right)
 T_{2,\mu^{-1} q_1 q_2 x}
 \right)
\end{align}
\begin{align}
 &
 f_{22}\left(\frac{y}{x}\right) T_{2,x} T_{2,y}
 - f_{22}\left(\frac{x}{y}\right) T_{2,y} T_{2,x}
 \nonumber \\
 & \hspace{4em}
 =
 - \frac{(1-q_1)(1-q_2)}{1-q_1q_2}
 \left(
  \delta \left( q_1 q_2 \frac{y}{x} \right) T_{1,\mu q_1^{-1} q_2^{-1} x}
 - \delta \left( q_1^{-1} q_2^{-1} \frac{y}{x} \right) T_{1,\mu x}
 \right)
 \nonumber \\[.5em]
 & \hspace{6em}
 - \frac{(1-q_1)(1-q_2)(1-q_1^2 q_2^2)(1-q_1^3 q_2)}
        {(1-q_1q_2)(1-q_1^2 q_2)(1-q_1^3 q_2^2)}
 %\msS_2 (q_1 q_2^2)
 \left(
 \delta \left( q_1^3 q_2^2 \frac{y}{x} \right)
 - \delta \left( q_1^{-3} q_2^{-2} \frac{y}{x} \right)
 \right)
\end{align}
where the $f$-factor is the contribution from the $\sY$-operator OPE
\begin{align}
 f_{ij}(x)
 & =
 \exp
 \left(
 \sum_{p=1}^\infty
 (1 - q_1^p)(1 - q_2^{d_j p}) \tilde{c}_{ij}^{[-p]} \, x^{p}
 \right)
 \, .
\end{align}
These OPEs define the algebraic relation of $\mu$-deformed W($BC_2$)-algebra, which is consistent with the construction given by~\cite{Bouwknegt:1998da} and~\cite{Frenkel:1997} in the classical limit.

\subsection{$B_r$ quiver}\label{sec:Br}

We consider $B_r$ quiver which consists of $r$ nodes with $d_i=2$ for $i = 1,\ldots, r-1$ and $d_r = 1$.
In this case the local iWeyl reflection is given by
\begin{align}
 \sY_{i,x}
 &
 \ \longrightarrow \
 \frac{\sY_{i-1,\mu_{i-1 \to i}q_1^{-2}q_2^{-1}x}
       \sY_{i+1,\mu_{i\to i+1}^{-1}x}}
      {\sY_{i,q_1^{-2}q_2^{-1}x}}
 \qquad
 (i = 1,\ldots,r-2)
 \\
 \sY_{r-1,x}
 &
 \ \longrightarrow \
 \frac{\sY_{r-2,\mu_{r-2 \to r-1} q_1^{-2} q_2^{-1} x}
       \sY_{r,\mu_{r-1 \to r}^{-1} x}
       \sY_{r,\mu_{r-1 \to r}^{-1} q_1^{-1} x}}
      {\sY_{r-1,q_1^{-2}q_2^{-1}x}}
 \\
 \sY_{r,x}
 &
 \ \longrightarrow \
 \frac{\sY_{r-1,\mu_{r-1 \to r} q_1^{-1} q_2^{-1} x}}
      {\sY_{r,q_1^{-1} q_2^{-1} x}}
\end{align}
where we put $\sY_{0,x}=1$.
Introduce the fields
\begin{align}
% \Lambda_{1,x}
% & = \sY_{1,x}
% \, , \\
 \Lambda_{i,x}
 & =
 \frac{\sY_{i,\mu_i^{-1} x}}{\sY_{i-1,\mu_{i-1}^{-1} q_1^{-2}q_2^{-1}x}}
 \qquad (i = 1,\ldots,r-1)
 \, , \\
 \Lambda_{r,x}
 & =
 \frac{\sY_{r,\mu_r^{-1}x} \sY_{r,\mu_r^{-1}q_2^{-1}x}}
      {\sY_{r-1,\mu_{r-1}^{-1}q_1^{-2}q_2^{-1}x}}
 \, , \\
 \Lambda_{r+1,x}
 & =
 \frac{(1 + q_1)(1 - q_1 q_2)}{1 - q_1^2 q_2}
 \frac{\sY_{r,\mu_r^{-1}x}}{\sY_{r,\mu_r^{-1}q_1^{-2}q_2^{-1}x}}
 \, , \\
 \Lambda_{r+2,x}
 & =
 \frac{\sY_{r-1,\mu_{r-1}^{-1} q_1^{-1} q_2^{-1} x}}
      {\sY_{r,\mu_{r}^{-1} q_1^{-1} q_2^{-1} x} \sY_{r,\mu_{r}^{-1} q_1^{-2} q_2^{-1} x}} 
 \, , \\  
 \Lambda_{2r+2-i,x}
 & =
 \frac{\sY_{i-1,\mu_{i-1}^{-1} q_1^{-3} q_2^{-2} x}}
      {\sY_{i,\mu_{i}^{-1} q_1^{-3} q_2^{-2} x}}
 \qquad (i = 1,\ldots,r-1)
\end{align}
where we parametrize the mass parameters
\begin{align}
 \mu_i
 := \mu_{1 \to 2} \mu_{2 \to 3} \cdots \mu_{i-1 \to i}
 = \prod_{j=1}^{i-1} \mu_{j \to j+1}
 \label{eq:mass_prod}
\end{align}
with $\mu_1 = 1$.
Then the fundamental $qq$-character is given by~\cite{Frenkel:1996,Frenkel:1997}
\begin{align}
 T_{1,x}
 & =
 \sum_{i=1}^{2r+1}
 \Lambda_{i,x}
 \, ,
\end{align}
which corresponds to the $(2r+1)$-dimensional vector representation of $SO(2r+1)$.

For example, we have three $qq$-characters for $B_3$ quiver,
\begin{align}
 T_{1,x}
 & =
 \sY_{1,x}
 + \frac{\sY_{2,\mu_2^{-1} x}}{\sY_{1,q_1^{-2} q_2^{-1} x}}
 + \frac{\sY_{3,\mu_3^{-1} x} \sY_{3,\mu_3^{-1} q_1^{-1} x}}
        {\sY_{2,\mu_2^{-1} q_1^{-2} q_2^{-1} x}}
 + \msS(q_1) \frac{\sY_{3,\mu_3^{-1} x}}{\sY_{3,\mu_3^{-1} q_1^{-2} q_2^{-1} x}}
 \nonumber \\
 & \quad
 + \frac{\sY_{2,\mu_2^{-1} q_1^{-1} q_2^{-1} x}}
        {\sY_{3,\mu_3^{-1} q_1^{-1} q_2^{-1} x} \sY_{3,\mu_3^{-1} q_1^{-2} q_2^{-1} x}}
 + \frac{\sY_{1,q_1^{-3} q_2^{-2} x}}{\sY_{2,\mu_2^{-1} q_1^{-3} q_2^{-2} x}}
 + \frac{1}{\sY_{1,q_1^{-5} q_2^{-3} x}} 
 \, ,
\end{align}
\begin{align}
 T_{2,\mu_2^{-1} x}
 & =
 \sY_{2,\mu_2^{-1} x}
 +
 \frac{\sY_{1,q_1^{-2} q_2^{-1} x}
       \sY_{3,\mu_3^{-1}} \sY_{3,\mu_3^{-1} q_1^{-1} x}}
      {\sY_{2,\mu_2^{-1} q_1^{-2} q_2^{-1} x}}
 +
 \frac{\sY_{3,\mu_3^{-1} x} \sY_{3,\mu_3^{-1} q_1^{-1} x}}
      {\sY_{1,q_1^{-4} q_2^{-2} x}}
 +
 \msS(q_1)
 \frac{\sY_{1,q_1^{-2} q_2^{-1} x} \sY_{3,\mu_3^{-1} x}}
      {\sY_{3,\mu_3^{-1} q_1^{-2} q_2^{-1} x}}
 \nonumber \\[.5em]
 & \quad
 +
 \msS(q_1)
 \frac{\sY_{3,\mu_3^{-1} x} \sY_{3,\mu_3^{-1} q_1^{-3} q_2^{-1} x}}
      {\sY_{2,\mu_2^{-1} q_1^{-4} q_2^{-2} x}}
 +
 \frac{\sY_{2,\mu_2^{-1} q_1^{-1} q_2^{-1} x} \sY_{2,\mu_2^{-1} q_1^{-2} q_2^{-1} x}}
      {\sY_{1,q_1^{-4} q_2^{-2} x}
       \sY_{3,\mu_3^{-1} q_1^{-1} q_2^{-1} x} \sY_{3,\mu_3^{-1} q_1^{-2} q_2^{-1} x}}
 +
 \frac{\sY_{1,q_1^{-2} q_2^{-1} x} \sY_{1,q_1^{-3} q_2^{-2} x}}
      {\sY_{2,\mu_2^{-1} q_1^{-3} q_2^{-2} x}}
 \nonumber \\[.5em]
 & \quad
 +
 \msS(q_1)
 \frac{\sY_{2,\mu_2^{-1} q_1^{-1} q_2^{-1} x} \sY_{3,\mu_3^{-1} x}}{\sY_{1,q_1^{-4} q_2^{-2} x} \sY_{3,\mu_3^{-1} q_1^{-2} q_2^{-1} x}}
 +
 \frac{\sY_{1,q_1^{-2} q_2^{-1} x} \sY_{2,\mu_2^{-1} q_1^{-1} q_2^{-1} x}}
      {\sY_{3,\mu_3^{-1} q_1^{-1} q_2^{-1} x} \sY_{3,\mu_3^{-1} q_1^{-2} q_2^{-1} x}}
  \nonumber \\[.5em]
 & \quad
 +
% \msS_{q_2} \msS_{q_1q_2^3}
 \msS(q_1) \msS(q_1^3 q_2)
 \frac{\sY_{3,\mu_3^{-1} x}}{\sY_{3,\mu_3^{-1} q_1^{-4} q_2^{-2} x}}
 +
% \msS_{q_2}^{(2)}
 \msS_2(q_1)
 \frac{\sY_{2,\mu_2^{-1} q_1^{-1} q_2^{-1} x} \sY_{3,\mu_3^{-1} q_1^{-3} q_2^{-1} x}}
      {\sY_{2,\mu_2^{-1} q_1^{-4} q_2^{-2} x} \sY_{3,\mu_3^{-1} q_1^{-1} q_2^{-1} x}}
 +
% \msS_{q_2^{-1}}^{(2)}
 \msS_2(q_1^{-1})
 \frac{\sY_{1,q_1^{-3} q_2^{-2} x} \sY_{2,\mu_2^{-1} q_1^{-2} q_2^{-1} x}}
      {\sY_{1,q_1^{-4} q_2^{-2} x} \sY_{2,\mu_2^{-1} q_1^{-3} q_2^{-2} x}}
 \nonumber \\[.5em]
 & \quad
 +
% \msS_{q_2}^{(2)}
 \msS_2(q_1 q_2)
 \frac{\sY_{1,q_1^{-2} q_2^{-1} x}}{\sY_{1,q_1^{-5} q_2^{-3} x}}
 +
 \msS(q_1)
 \frac{\sY_{2,\mu_2^{-1} q_1^{-1} q_2^{-1} x}}
      {\sY_{3,\mu_3^{-1} q_1^{-1} q_2^{-1} x} \sY_{3,\mu_3^{-1} q_1^{-4} q_2^{-2} x}}
 +
 \frac{\sY_{1,q_1^{-3} q_2^{-2} x}
       \sY_{3,\mu_3^{-1} q_1^{-2} q_2^{-1} x} \sY_{3,\mu_3^{-1} q_1^{-3} q_2^{-1} x}}
      {\sY_{2,\mu_2^{-1} q_1^{-3} q_2^{-2} x} \sY_{2,\mu_2^{-1} q_1^{-4} q_2^{-2} x}}
 \nonumber \\[.5em]
 & \quad
 +
 \frac{\sY_{2,\mu_2^{-1} q_1^{-2} q_2^{-1} x}}
      {\sY_{1,q_1^{-4} q_2^{-2} x} \sY_{1,q_1^{-5} q_2^{-3} x}}
 +
 \msS(q_1)
 \frac{\sY_{1,q_1^{-3} q_2^{-2} x} \sY_{3,\mu_3^{-1} q_1^{-2} q_2^{-1} x}}
      {\sY_{2,\mu_2^{-1} q_1^{-3} q_2^{-2} x} \sY_{3,\mu_3^{-1} q_1^{-4} q_2^{-2} x}}
 +
 \frac{\sY_{3,\mu_3^{-1} q_1^{-2} q_2^{-1} x} \sY_{3,\mu_3^{-1} q_1^{-3} q_2^{-1} x}}
      {\sY_{1,q_1^{-5} q_2^{-3} x} \sY_{2,\mu_2^{-1} q_1^{-4} q_2^{-2} x}}
 \nonumber \\[.5em]
 & \quad
 +
 \frac{\sY_{1,q_1^{-3} q_2^{-2} x}}
      {\sY_{3,\mu_3^{-1} q_1^{-3} q_2^{-2} x} \sY_{3,\mu_3^{-1} q_1^{-4} q_2^{-2} x}}
 +
 \msS(q_1)
 \frac{\sY_{3,\mu_3^{-1} q_1^{-2} q_2^{-1} x}}
      {\sY_{1,q_1^{-5} q_2^{-3} x} \sY_{3,\mu_3^{-1} q_1^{-4} q_2^{-2} x}}
 \nonumber \\[.5em]
 & \quad
 +
 \frac{\sY_{2,\mu_2^{-1} q_1^{-3} q_2^{-2} x}}
      {\sY_{1,q_1^{-5} q_2^{-3} x}
       \sY_{3,\mu_3^{-1} q_1^{-3} q_2^{-2} x} \sY_{3,\mu_3^{-1} q_1^{-4} q_2^{-2} x}}
 +
 \frac{1}{\sY_{2,\mu_2^{-1} q_1^{-5} q_2^{-3} x}}
 \, ,
\end{align}
\begin{align}
 T_{3,\mu_{3}^{-1} x}
 & =
 \sY_{3,\mu_3^{-1} x}
 + \frac{\sY_{2,\mu_{2}^{-1} q_1^{-1} q_2^{-1} x}}
        {\sY_{3,\mu_3^{-1} q_1^{-1} q_2^{-1} x}}
 + \frac{\sY_{1,q_1^{-3} q_2^{-2} x} \sY_{3,\mu_{3}^{-1} q_1^{-2} q_2^{-1} x}}
        {\sY_{2,\mu_{2}^{-1} q_1^{-3} q_2^{-2} x}}
 + \frac{\sY_{3,\mu_3^{-1} q_1^{-2} q_2^{-1} x}}{\sY_{1,q_1^{-5} q_2^{-3} x}}
 \nonumber \\[.5em]
 & \quad
 + \frac{\sY_{1,q_1^{-3} q_2^{-2} x}}{\sY_{3,\mu_3^{-1} q_1^{-3} q_2^{-2} x}}
 + \frac{\sY_{2,\mu_2^{-1} q_1^{-3} q_2^{-2} x}}
        {\sY_{1,q_1^{-5} q_2^{-3} x} \sY_{3,\mu_{3}^{-1} q_1^{-3} q_2^{-2} x}}
 + \frac{\sY_{3,\mu_3^{-1} q_1^{-4} q_2^{-2} x}}{\sY_{2,\mu_2^{-1} q_1^{-5} q_2^{-3} x}}
 + \frac{1}{\sY_{3,\mu_3^{-1} q_1^{-5} q_2^{-3} x}}
 \, . 
\end{align}
They correspond to \textbf{7} (vector), \textbf{21} (adjoint), and \textbf{8} (spinor) representations, respectively.
There are several $\msS$-factors in the expressions which are peculiar to the $qq$-character.

\subsection{$C_r$ quiver}\label{sec:Cr}

The $C_r$ quiver consists of $r$ nodes with $d_i=1$ for $i = 1,\ldots, r-1$ and $d_r = 2$.
The local iWeyl reflection is
\begin{align}
 \sY_{i,x}
 &
 \ \longrightarrow \
 \frac{\sY_{i-1,\mu_{i-1 \to i} q_1^{-1} q_2^{-1} x}
       \sY_{i+1,\mu_{i \to i+1}^{-1} x}}
      {\sY_{i,q_1^{-1} q_2^{-1} x}}
 \qquad
 (i = 1, \ldots, r-1)
 \\
 \sY_{r,x}
 &
 \ \longrightarrow \
 \frac{\sY_{r-1,\mu_{r-1 \to r} q_1^{-1} q_2^{-1} x}
       \sY_{r-1,\mu_{r-1 \to r} q_1^{-2} q_2^{-1} x}}
      {\sY_{r,q_1^{-2} q_2^{-1} x}}
 \, .
\end{align}
Introducing the fields
\begin{align}
% \Lambda_{1,x}
% & = \sY_{1,x}
% \, , \\
 \Lambda_{i,x}
 & =
 \frac{\sY_{i,\mu_i^{-1} x}}{\sY_{i-1,\mu_{i-1}^{-1} q_1^{-1}q_2^{-1}x}}
 \qquad (i = 1,\ldots,r)
 \, , \\
 \Lambda_{r+1,x}
 & =
 \frac{\sY_{r-1,\mu_{r-1}^{-1} q_1^{-2} q_2^{-1} x}}
      {\sY_{r,\mu_{r}^{-1} q_1^{-2} q_2^{-1} x}} 
 \, , \\
 \Lambda_{2r+1-i,x}
 & =
 \frac{\sY_{i-1,\mu_{i-1}^{-1} q_1^{-3} q_2^{-2} x}}
      {\sY_{i,\mu_{i}^{-1} q_1^{-3} q_2^{-2} x}}
 \qquad (i = 1,\ldots,r-1)
 \, ,
\end{align}
the fundamental $qq$-character is given by~\cite{Frenkel:1996,Frenkel:1997}
\begin{align}
 T_{1,x}
 & =
 \sum_{i=1}^{2r}
 \Lambda_{i,x}
\end{align}
which corresponds to the $2r$-dimensional representation of $Sp(r)$.
Here we use the same notation for the mass parameter as before \eqref{eq:mass_prod}.

The $qq$-characters for $C_3$ quiver are explicitly given as follows:
\begin{align}
 T_{1,x}
 & =
 \sY_{1,x}
 +
 \frac{\sY_{2,\mu_2^{-1} x}}{\sY_{1,q_1^{-1} q_2^{-1} x}}
 +
 \frac{\sY_{3,\mu_3^{-1} x}}{\sY_{2,\mu_{2}^{-1} q_1^{-1} q_2^{-1} x}}
 +
 \frac{\sY_{2,\mu_2^{-1} q_1^{-2} q_2^{-1} x}}
      {\sY_{3,\mu_3^{-1} q_1^{-2} q_2^{-1} x}}
 +
 \frac{\sY_{1,q_1^{-3} q_2^{-2} x}}{\sY_{2,\mu_2^{-1} q_1^{-3} q_2^{-2} x}}
 +
 \frac{1}{\sY_{1,q_1^{-4} q_2^{-3} x}}
 \, ,
\end{align}
\begin{align}
 T_{2,\mu_2^{-1} x}
 & =
 \sY_{2,\mu_2^{-1} x}
 +
 \frac{\sY_{1,q_1^{-1} q_2^{-1} x} \sY_{3,\mu_3^{-1} x}}
      {\sY_{2,\mu_2^{-1} q_1^{-1} q_2^{-1} x}}
 +
 \frac{\sY_{3,\mu_3^{-1} x}}{\sY_{1,q_1^{-2} q_2^{-2} x}}
 +
 \frac{\sY_{1,q_1^{-1} q_2^{-1} x} \sY_{2,\mu_2^{-1} q_1^{-2} q_2^{-1} x}}
      {\sY_{3,\mu_3^{-1} q_1^{-2} q_2^{-1} x}}
 +
 \frac{\sY_{2,\mu_2^{-1} q_1^{-1} q_2^{-1} x} \sY_{2,\mu_2^{-1} q_1^{-2} q_2^{-1} x}}
      {\sY_{1,q_1^{-2} q_2^{-2} x} \sY_{3,\mu_3^{-1} q_1^{-2} q_2^{-1} x}}
 \nonumber \\[.5em]
 & \quad
 +
 \frac{\sY_{1,q_1^{-1} q_2^{-1} x} \sY_{1,q_1^{-3} q_2^{-2} x}}
      {\sY_{2,\mu_2^{-1} q_1^{-3} q_2^{-2} x}}
 +
 \msS(q_1)
 \frac{\sY_{1,q_1^{-3} q_2^{-2} x} \sY_{2,\mu_2^{-1} q_1^{-1} q_2^{-1} x}}
      {\sY_{1,q_1^{-2} q_2^{-2} x} \sY_{2,\mu_2^{-1} q_1^{-3} q_2^{-2} x}}
 +
 \msS(q_1^2 q_2)
 \frac{\sY_{1,q_1^{-1} q_2^{-1} x}}{\sY_{1,q_1^{-4} q_2^{-3} x}}
 \nonumber \\[.5em]
 & \quad
 +
 \frac{\sY_{1,q_1^{-3} q_2^{-2} x} \sY_{3,\mu_3^{-1} q_1^{-1} q_2^{-1} x}}
      {\sY_{2,\mu_2^{-1} q_1^{-2} q_2^{-2} x} \sY_{2,\mu_2^{-1} q_1^{-3} q_2^{-2} x}}
 +
 \frac{\sY_{3,\mu_3^{-1} q_1^{-1} q_2^{-1} x}}
      {\sY_{1,q_1^{-4} q_2^{-3} x} \sY_{2,\mu_2^{-1} q_1^{-2} q_2^{-2} x}}
 +
 \frac{\sY_{1,q_1^{-3} q_2^{-2} x}}{\sY_{3,\mu_3^{-1} q_1^{-3} q_2^{-2} x}}
 +
 \frac{\sY_{2,\mu_2^{-1} q_1^{-1} q_2^{-1} x}}{\sY_{1,q_1^{-2} q_2^{-2} x} \sY_{1,q_1^{-4} q_2^{-3} x}}
 \nonumber \\[.5em]
 & \quad
 +
 \frac{\sY_{2,\mu_2^{-1} q_1^{-3} q_2^{-2} x}}
      {\sY_{1,q_1^{-4} q_2^{-3} x} \sY_{3,\mu_3^{-1} q_1^{-3} q_2^{-2} x}}
 +
 \frac{1}{\sY_{2,\mu_2^{-1} q_1^{-4} q_2^{-3} x}}
 \, ,
\end{align}
\begin{align}
 T_{3,\mu_3^{-1} x}
 & =
 \sY_{3,\mu_3^{-1} x}
 +
 \frac{\sY_{2,\mu_2^{-1} q_1^{-1} q_2^{-1} x} \sY_{2,\mu_2^{-1} q_1^{-2} q_2^{-1} x}}
      {\sY_{3,\mu_3^{-1} q_1^{-2} q_2^{-1} x}}
 +
 \msS(q_1)
 \frac{\sY_{1,q_1^{-3} q_2^{-2} x} \sY_{2,\mu_2^{-1} q_1^{-1} q_2^{-1} x}}
      {\sY_{2,\mu_2^{-1} q_1^{-3} q_2^{-2} x}}
 +
 \msS(q_1)
 \frac{\sY_{2,\mu_2^{-1} q_1^{-1} q_2^{-1} x}}
      {\sY_{1,q_1^{-4} q_2^{-3} x}}
 \nonumber \\[.5em]
 & \quad
 +
 \frac{\sY_{1,q_1^{-2} q_2^{-2} x} \sY_{1,q_1^{-3} q_2^{-2} x}
       \sY_{3,\mu_3^{-1} q_1^{-1} q_2^{-1} x}}
      {\sY_{2,\mu_2^{-1} q_1^{-2} q_2^{-2} x} \sY_{2,\mu_2^{-1} q_1^{-3} q_2^{-2} x}}
 +
 \msS(q_1)
 \frac{\sY_{1,q_1^{-2} q_2^{-2} x} \sY_{3,\mu_3^{-1} q_1^{-1} q_2^{-1} x}}
      {\sY_{1,q_1^{-4} q_2^{-3} x} \sY_{2,\mu_2^{-1} q_1^{-2} q_2^{-2} x}}
 +
 \frac{\sY_{1,q_1^{-2} q_2^{-2} x} \sY_{1,q_1^{-3} q_2^{-2} x}}
      {\sY_{3,\mu_3^{-1} q_1^{-3} q_2^{-2} x}}
 \nonumber \\[.5em]
 & \quad
 +
 \msS(q_1)
 \frac{\sY_{1,q_1^{-2} q_2^{-2} x} \sY_{2,\mu_2^{-1} q_1^{-3} q_2^{-2} x}}
      {\sY_{1,q_1^{-4} q_2^{-3} x} \sY_{3,\mu_3^{-1} q_1^{-3} q_2^{-2} x}}
 +
 \frac{\sY_{3,\mu_3^{-1} q_1^{-1} q_2^{-1} x}}
      {\sY_{1,q_1^{-3} q_2^{-3} x} \sY_{1,q_1^{-4} q_2^{-3} x}}
 +
 \frac{\sY_{2,\mu_2^{-1} q_1^{-2} q_2^{-2} x} \sY_{2,\mu_2^{-1} q_1^{-3} q_2^{-2} x}}
      {\sY_{1,q_1^{-3} q_2^{-3} x} \sY_{1,q_1^{-4} q_2^{-3} x} \sY_{3,\mu_3^{-1} q_1^{-3} q_2^{-2} x}}
 \nonumber \\[.5em]
 & \quad
 +
 \msS(q_1)
 \frac{\sY_{1,q_1^{-2} q_2^{-2} x}}{\sY_{2,\mu_2^{-1} q_1^{-4} q_2^{-3} x}}
 +
 \msS(q_1)
 \frac{\sY_{2,\mu_2^{-1} q_1^{-2} q_2^{-2} x}}
      {\sY_{1,q_1^{-3} q_2^{-3} x} \sY_{2,\mu_2^{-1} q_1^{-4} q_2^{-3} x}}
 +
 \frac{\sY_{3,\mu_3^{-1} q_1^{-2} q_2^{-2} x}}
      {\sY_{2,\mu_2^{-1} q_1^{-3} q_2^{-3} x} \sY_{2,\mu_2^{-1} q_1^{-4} q_2^{-3} x}}
 +
 \frac{1}{\sY_{3,\mu_3^{-1} q_1^{-4} q_2^{-3} x}}
 \, .
\end{align}
They correspond to the 6, 15, and 14 dimensional representations of $Sp(3)$.

\subsection{Affine fractional quiver}
%\subsubsection{4-1 quiver}

We consider the affine fractional quiver:
\begin{center}
 \begin{tikzpicture}[thick,baseline=0]

  \draw (0,.15) circle [radius = .67em] node (1) {$4$};
  
  \draw (1,.15) circle [radius = .67em] node (2) {$1$};
  
  \draw (1) -- (2);

  \node at (0,-.5) {$1$};
  \node at (1,-.5) {$2$};  

  \node at (-.5,-.45) [left] {node:};
  
\end{tikzpicture}
\end{center}
which corresponds to the twisted affine Lie algebra $A_1^{(2)}$.
In the standard notation, the quiver--Dynkin diagram is given by
\begin{tikzpicture}[thick,baseline=0]

% \draw [double,double distance = 1.pt] (0,.07) -- ++(1,0); %(1) -- (2);
% \draw [double,double distance = 1.pt] (0,.23) -- ++(1,0); %(1) -- (2); 
 \draw (0,.06) -- ++(1,0);
 \draw (0,.12) -- ++(1,0);
 \draw (0,.18) -- ++(1,0);
 \draw (0,.24) -- ++(1,0); 
 
 \filldraw[fill = white] (0,.15) circle [radius = .35em] node (1) {};

 \filldraw[fill = white] (1,.15) circle [radius = .35em] node (2) {};

 \draw (.4,.35) -- (.6,.15) -- (.4,-.05);
 
\end{tikzpicture}.
The mass-deformed Cartan matrix in this case is
\begin{align}
 (c_{ij})
 & =
 \begin{pmatrix}
  1 + q_1^{-4} q_2^{-1} & - \mu^{-1}  \\
  - \mu q_1^{-1}q_2^{-1} (1 + q_1^{-1} + q_1^{-2} + q_1^{-3})
  & 1 + q_1^{-1} q_2^{-1}
 \end{pmatrix}
 \ \stackrel{(c_{ij}^{[0]})}{\longrightarrow} \
 \begin{pmatrix}
  2 & -1 \\ -4 & 2
 \end{pmatrix}
 \, .
\end{align}
Here the mass parameter is defined in the same way as \eqref{eq:BC_mass}, and the 0-th Adams operation $(c_{ij}^{[0]})$ provides the ordinary Cartan matrix \eqref{eq:Cartan_0th}.
The determinant is given by
\begin{align}
 \det (c_{ij})
 & =
 1 + q_1^{-5} q_2^{-2}
 - q_1^{-2} q_2^{-1} (1 + q_1^{-1})
 \ \stackrel{(c_{ij}^{[0]})}{\longrightarrow} \
 0
 \, .
\end{align}
Thus the Cartan matrix $(c_{ij}^{[0]})$ is not invertible.
We remark that the determinant does not depend on the mass parameter $\mu$.

The iWeyl reflection associated with this quiver is given by
\begin{align}
 \sY_{1,x}
 & \ \longrightarrow \
 \frac{q_1^{-4} \fq_1}{\sY_{1,q_1^{-1} q_2^{-4} x}}
 \sY_{2,\mu^{-1} q_2^{-3} x} \sY_{2,\mu_{e}^{-1} q_2^{-2} x}
 \sY_{2,\mu^{-1} q_2^{-1} x} \sY_{2,\mu_e^{-1} x} \\
 \sY_{2,x}
 & \ \longrightarrow \
 \frac{q_1^{-1} \fq_2}{\sY_{2,q_1^{-1} q_2^{-1} x}}
 \sY_{1,\mu_e q_1^{-1} q_2^{-1} x}
 \, .
\end{align}
In this case we need to assign the coupling constant $\fq_i$ and the factor $q_1$ to each reflection, since the Cartan matrix $(c_{ij}^{[0]})$ is not invertible.
The $\sY$-operator zero mode cannot absorb them.
Then the fundamental $qq$-characters are generated as follows,
\begin{align}
 T_{1,x}
 & =
 \sY_{1,x}
 +
 q_1^{-4} \fq_1 \,
 \frac{
 \sY_{2,\mu^{-1} q_1^{-3}x} \sY_{2,\mu^{-1} q_1^{-2}x}
 \sY_{2,\mu^{-1} q_1^{-1}x} \sY_{2,\mu^{-1}x}
 }{\sY_{1,q_1^{-4} q_2^{-1} x}}
 \nonumber \\
 & \quad
 +
 q_1^{-5} \fq_1 \fq_2 \,
% \msS(q_1) \msS(q_1^2) \msS(q_1^3)
 \msS_3(q_1)
 \frac{\sY_{2,\mu^{-1}x} \sY_{2,\mu^{-1} q_1^{-1}x}
       \sY_{2,\mu^{-1} q_1^{-2}x}}
      {\sY_{2,\mu^{-1} q_1^{-4} q_2^{-1} x}}
% \nonumber \\
% & \quad
 +
 q_1^{-5} \fq_1 \fq_2 \,
% \msS(q_1)^2 \msS(q_1^2)^2 \msS(q_1^3)
 \msS_3(q_1)^2 
 \frac{\sY_{1,q_1^{-3} q_2^{-1}x} \sY_{2,\mu^{-1}x}
       \sY_{2,\mu^{-1} q_1^{-1}x}}
      {\sY_{2,\mu^{-1} q_1^{-3} q_2^{-1}x}
       \sY_{2,\mu^{-1} q_1^{-4} q_2^{-1}x}}
 \nonumber \\
 & \quad + \cdots
 \, , \\
 T_{2,x}
 & =
 \sY_{2,x}
 + q_1^{-1} \fq_2 \,
 \frac{\sY_{1,\mu q_1^{-1} q_2^{-1}x}}{\sY_{2,q_1^{-1} q_2^{-1}x}}
 + q_1^{-5} \fq_1 \fq_2 \,
 \frac{\sY_{2,q_1^{-2} q_2^{-1}x}
       \sY_{2,q_1^{-3} q_2^{-1}x}
       \sY_{2,q_1^{-4} q_2^{-1}x}}
      {\sY_{1,\mu q_1^{-5} q_2^{-2}x}}
 \nonumber \\
 & \quad
 + q_1^{-6} \fq_1 \fq_2^2 \,
% \msS(q_1) \msS(q_1^2)
 \msS_2(q_1)
 \frac{\sY_{2,q_1^{-2} q_2^{-1}x} \sY_{2,q_1^{-3} q_2^{-1}x}}
      {\sY_{2,q_1^{-5} q_2^{-2}x}}
 + q_1^{-6} \fq_1 \fq_2^2 \,
% \msS(q_1)^2 \msS(q_1^2)
 \msS_2(q_1)^2 
 \frac{\sY_{2,q_1^{-2} q_2^{-1}x} \sY_{1,\mu q_1^{-4} q_2^{-2}x}}
      {\sY_{2,q_1^{-4} q_2^{-2}x} \sY_{2,q_1^{-5} q_2^{-2}x}}
 \nonumber \\
 & \quad
 + \cdots
 \, .
\end{align}
These $qq$-characters commute with the screening charge $\left[T_{i,x}, \sS_{j,x'}\right]=0$, and involve infinitely many monomials of the $\sY$-operators, since the corresponding fundamental representations are infinite-dimensional.

\subsection{Hyperbolic fractional quiver}

We then consider the hyperbolic fractional quiver:
\begin{center}
 \begin{tikzpicture}[thick,baseline=0]

  \draw (0,.15) circle [radius = .67em] node (1) {$3$};
  
  \draw (1,.15) circle [radius = .67em] node (2) {$2$};
  
  \draw (1) -- (2);

  \node at (0,-.5) {$1$};
  \node at (1,-.5) {$2$};

  \node at (-.5,-.45) [left] {node:};  
 
\end{tikzpicture}
\end{center}
which is characterized by the mass-deformed Cartan matrix
\begin{align}
 (c_{ij})
 & =
 \begin{pmatrix}
  1 + q_1^{-3} q_2^{-1} & - \mu^{-1} (1 + q_1^{-1}) \\
  - \mu q_1^{-1}q_2^{-1} (1 + q_1^{-1} + q_1^{-2}) 
  & 1 + q_1^{-2} q_2^{-1}
 \end{pmatrix}
 \ \stackrel{(c_{ij}^{[0]})}{\longrightarrow} \
 \begin{pmatrix}
  2 & -2 \\ -3 & 2
 \end{pmatrix}
 \, .
\end{align}
The mass parameter is defined in the same way as \eqref{eq:BC_mass} as well.
The determinant is given by
\begin{align}
 \det (c_{ij})
 & =
 1 + q_1^{-5} q_2^{-2}
 - q_1^{-1} q_2^{-1} (1 + q_1^{-1} + q_1^{-2} + q_1^{-3})
 \ \stackrel{(c_{ij}^{[0]})}{\longrightarrow} \
 - 2
 \, .
\end{align}
Since the determinant of the Cartan matrix $(c_{ij}^{[0]})$ is negative, it is classified to the hyperbolic quiver.

The iWeyl reflection is given by
\begin{align}
 \sY_{1,x}
 & \ \longrightarrow \
 \frac{1}{\sY_{1,q_1^{-3} q_2^{-1} x}}
 \sY_{2,\mu^{-1} q_1^{-2} x} \sY_{2,\mu^{-1} q_1^{-1} x} \sY_{2,\mu^{-1} x}
 \, ,\\
 \sY_{2,x}
 & \ \longrightarrow \
 \frac{1}{\sY_{2,q_1^{-2} q_2^{-1} x}}
 \sY_{1,\mu q_1^{-2} q_2^{-1} x} \sY_{1,\mu q_1^{-1} q_2^{-1} x}
 \, ,
\end{align}
which generate the fundamental $qq$-characters
\begin{align}
 T_{1,x}
 & =
 \sY_{1,x}
 + \frac{\sY_{2,\mu^{-1}x} \sY_{2,\mu^{-1} q_1^{-1}x} \sY_{2,\mu^{-1} q_1^{-2}}x}{\sY_{1,q_1^{-3}q_2^{-1}x}}
 \nonumber \\
 & \quad
 +
 \msS_2(q_1) \msS_2(q_1^{-1})
 \frac{\sY_{1,q_1^{-2}q_2^{-1}x} \sY_{2,\mu^{-1}x} \sY_{2,\mu^{-1} q_1^{-2}x}}
      {\sY_{2,\mu^{-1}q_1^{-3}q_2^{-1}x}}
 \nonumber \\
 & \quad
 +
 \msS_2(q_1) \msS_2(q_1^{2})
 \frac{\sY_{1,q_1^{-4}q_2^{-1}x}\sY_{2,\mu^{-1}} \sY_{2,\mu^{-1} q_1^{-1}x}}
      {\sY_{2,\mu^{-1}q_1^{-4}q_2^{-1}x}}
 + \cdots 
 \, \\
 T_{2,x}
 & =
 \sY_{2,x}
 +
 \frac{\sY_{1,\mu q_1^{-1} q_2^{-1}x} \sY_{1,\mu q_1^{-2} q_2^{-1}x}}
      {\sY_{2,q_1^{-2} q_2^{-1}x}}
 \nonumber \\
 & \quad
 +
 \msS_3(q_1^{-1})
 \frac{\sY_{1,\mu q_1^{-2} q_2^{-1}x} \sY_{2,q_1^{-1} q_2^{-1}x} \sY_{2,q_1^{-2} q_2^{-1}x} \sY_{2,q_1^{-3} q_2^{-1}x}}
      {\sY_{1,\mu q_1^{-4} q_2^{-2}x} \sY_{2,q_1^{-2} q_2^{-1}x}}
 \nonumber \\
 & \quad
 +
 \msS_3(q_1)
 \frac{\sY_{1,\mu q_1^{-1} q_2^{-1}x} \sY_{2,q_1^{-2} q_2^{-1}x} \sY_{2,q_1^{-3} q_2^{-1}x} \sY_{2,q_1^{-4} q_2^{-1}x}}
      {\sY_{1,\mu q_1^{-5} q_2^{-2}x} \sY_{2,q_1^{-2} q_2^{-1}x}}
 + \cdots
 \, .
\end{align}
Since this quiver does not correspond to any finite dimensional Lie algebras, the $qq$-characters have infinitely many monomials of the $\sY$-operators, as well as the affine quiver.

%%%%%%%%%% References %%%%%%%%%%

%\bibliographystyle{utphysurl} \bibliography{wquiver}
\bibliographystyle{../wquiver/utphysurl}
\bibliography{../wquiver/wquiver} 

\end{document}